# Mentorship resources and citation-elite journal publication trajectories after training: Evidence from bioscience mentor-mentee networks

Hao Zeng, Yi Bu, Qingshan Zhou, and Hongkan Chen*

*Department of Information Management, Peking University, Beijing 100871, China*

*Correspondence concerning this article should be addressed to Hongkan Chen (chenhongkan@pku.edu.cn).

**Abstract:** Mentorship provides mentees with knowledge, collaborators, and reputation, but when do these resources support scholarship beyond training? Using 239,931 bioscience mentor–mentee pairs from the Academic Family Tree (1800–2020) linked to OpenAlex metadata, we examine this question through mentees' publication in citation-elite journals, identified by three source-level citation indicators. We distinguish training-period mentor and peer resources from post-graduation topic and coauthorship network continuity. Analyses combine trajectory comparisons, three propensity-score matching designs, PPML, two-part model and non-overlapping landmarks relating continuity in post-graduation years 1–3 to publication outcomes in years 4–10 and 6–20. Citation-elite journal publishing during mentorship is positively associated with later publishing, but correlates differ by career stage. Close mentor collaboration and larger peer group size characterize training-period publishing, whereas both are negatively associated with post-graduation publication probability and intensity. Early topic and network continuity are positively associated with later output in landmark analyses; across extended conventional windows, topic continuity becomes less favorable, consistent with later agenda broadening. Network continuity shows the most stable positive association with publication intensity. The findings suggest a stage-dependent resource-conversion process: training resources may facilitate entry into citation-elite journal publishing, whereas sustained output is associated with how intellectual and relational ties are retained or reconfigured.

## INTRODUCTION

As global competition in higher education intensifies and the academic labor market becomes increasingly saturated (Lauder & Brown, 2011; Lauder & Mayhew, 2020; Musselin, 2018), increasing attention has been directed toward how graduates sustain scholarly development and maintain research productivity after leaving the supervisory relationship in an increasingly competitive academic environment. As a core form of intellectual apprenticeship and a central gateway through which novice researchers enter the academic community (Weidman et al., 2001; Morita & Kobayashi, 2008), mentorship plays a pivotal role in shaping mentees' academic development and professional trajectories (Scandura, 1992; Wright & Wright, 1987). Consequently, contemporary discussions of graduate education and academic workforce development have increasingly examined whether mentorship can cultivate the capacity for sustained and independent scholarly production beyond the training period (Slaughter & Rhoades, 2004).

The resources support provided by mentors is widely regarded as a critical form of capital shaping doctoral socialization and early career development by prior research. Specifically, a mentor's scholarly reputation and academic capital confer symbolic advantages that enhance mentees' visibility and legitimacy within the field (Christopher Baker, 2021; Committee on Effective Mentoring in STEMM et al., 2019). Financial support and hands-on research guidance, meanwhile, constitute the material and cognitive foundations that enable mentees to complete their dissertations and generate early publications (Karakose et al., 2016; Kram, 1988; Y. Wang et al., 2026). In addition, being embedded in lab-mate or co-advisee collaboration networks offers access to knowledge spillovers, emotional support, and opportunities for tacit learning, further expanding the pool of academic resources available to mentees (Ruth et al., 2026; Ynalvez et al., 2017).

Though existing studies have demonstrated the short-term benefits of mentor support during doctoral training for young scholars to gain initial entry into the academic world (Bland et al., 2005), accumulate academic capital (Paglis et al., 2006) and enhance subsequent scholarly performance (Pinheiro et al., 2014). Other studies highlight potential constraints, such as unequal distribution of academic capital and dependence within hierarchical supervisory relationships (Long & Fox, 1995), the transmission of institutional pressures from supervisors to mentees (Gardner, 2010), and the risk that excessive reliance on mentors or highly homogeneous collaboration networks may constrain the development of distinctive research agendas and intellectual autonomy (Xing et al., 2025). These contrasting perspectives suggest that the long-term consequences of mentorship are neither uniform nor fully understood, and recent studies have begun to question whether the quantity or intensity of resources alone may be insufficient to explain heterogeneity in post-graduation research trajectories (Kadioglu & Valli, 2025; J. Li et al., 2025), since the effects of such resources may operate through the capabilities mentees develop during the mentorship. From a sociology of knowledge perspective, mentees develop research capabilities through the transformation of tacit knowledge acquired from engaging with academic resources into practical wisdom, such as cognitive schemas, strategic repertoires, and relational skills, and the progress is inherently gradual and may not be fully observable at primarily one point after graduation (Flyvbjerg & Sampson, 2011; Polanyi, 1967). Collectively, existing studies, which primarily focus on mentor-provided resources and evaluate outcomes within a limited post-graduation career window, remain insufficient to fully explain the factors influencing mentees' sustained scholarly development and long-term research productivity.

Against this backdrop, we employ a multi-stage empirical strategy to examine how supervisory experiences shape mentees' scholarly development across multiple career windows. Through descriptive heterogeneity analyses, propensity score matching, and regression analyses, we examine three research questions:

- RQ1: What is the relationship between citation-elite journal publication during the mentorship period and post-graduation outcomes?

- RQ2: What is the relationship between resource-related factors and citation-elite journal publication, and how does this relationship vary between the mentorship and post-graduation periods?
- RQ3: What is the relationship between continuity and citation-elite journal publication, and how does this relationship vary between the mentorship and post-graduation periods?

# RELATED WORK

*Mentorship and scholarly development*

Early conceptualizations of mentorship primarily focused on the functions that mentors provide to mentees. Kram (1985) characterized mentorship as a set of career-related and psychosocial supports offered within a master–apprentice relationship. Building on this perspective, Scandura (1992) further articulated mentorship as encompassing multiple dimensions, including career guidance, role modeling, and social support.

During the process where mentors foster deeper professional connections with mentees, this relationship provides sustained support (L. T. Eby & Dolan, 2015), operating through the transmission of disciplinary knowledge, research skills development, constructive feedback, role modeling, career guidance, access to collaborative and professional networks, and psychosocial support (Howard et al., 2006). Collectively, these functions have been associated with stronger research capabilities, as well as greater passion (McGee & Keller, 2007; Williams et al., 2016), and these benefits can translate into higher scholarly productivity (Steiner et al., 2002, 2004), bringing self-efficacy (Tenenbaum et al., 2001; Curtin et al., 2016), program satisfaction and clear professional identities and academic career aspirations (Paglis et al., 2006; Lunsford, 2012).

Despite its widely recognized benefits, the effects of mentorship are neither uniformly positive nor consistent across individuals. Eby et al. (2008, 2013) indicated considerable variation in mentoring outcomes across different types of mentoring relationships. Moreover, L. T. Eby et al. (2000) and Limeri et al. (2019) also noted that not all mentoring relationships are equally effective, and some may even encounter negative mentoring experiences, including mentor absence, misuse of power, interpersonal incompatibility, inadequate career or psychosocial support, misaligned expectations, and unequal treatment. While the entire relationship may not be harmful (L. T. Eby & McManus, 2004; Scandura, 1998), such experiences may coexist with otherwise supportive interactions, reflecting the complexity of prolonged mentor–mentee relationships (Cupach & Spitzberg, 1994; Kram, 1983).

These findings suggest that the role of mentorship in shaping scholarly development is heterogeneous: while some mentees benefit substantially from supportive relationships, others may experience limited or uneven gains. Recognizing that this heterogeneity may arise from both unequal access to mentoring resources and differences in capability development, this study distinguishes the resources available during mentorship from the capabilities mentees develop and deploy in their subsequent scholarly careers.

*Mentor guidance and capability formation*

Mentorship characterized by overall satisfaction, trust, effectiveness, and reciprocity provides mentees with both career-related and psychosocial support (L. T. D. T. Eby et al., 2013; L. T. Eby & Robertson, 2020). The acquisition of career-related resources—such as research skills, professional guidance, and access to academic networks—as well as psychosocial support, including encouragement, role modeling, and the development of professional identity in the mentorship, can facilitate mentees' academic and professional development (Committee on Effective Mentoring in STEMM et al., 2019). Mentor guidance takes multiple forms within doctoral supervision, encompassing not only direct instruction and feedback, but also collaborative practices such as coauthorship and joint research activities (Lee, 2008). In particular, collaboration with mentors unfolds through the development of supportive relationships, within which mentees engage in real research processes while benefiting from their mentors' expertise, reputation, and academic networks—helping them identify feasible research directions, navigate early-stage academic choices (Howard et al., 2006; Lankau & Scandura, 2002), and gain access to broader professional networks and visibility through association with established scholars (Kram, 1985, 1988; Lee, 2008). Such collaborative exposure may further allow mentees to acquire tacit knowledge of publication practices and enter high-impact venues through a "chaperone effect" (Sekara et al., 2018), while early coauthorship with high-status scientists may generate persistent advantages in subsequent academic careers (W. Li et al., 2019).

However, while such resource advantages facilitate early success, the long-term benefits of mentorship may depend on whether mentees develop intellectual independence (Lin et al., 2026; Ma et al., 2020). With the aim of fostering independent research capability, mentorship requires an appropriate combination of academic guidance and autonomy support that enables mentees to develop confidence and ownership of their research (Overall et al., 2011; Pol, 2025). More concretely, such capabilities encompass at least three interrelated dimensions. First, research autonomy, reflected in the capacity to independently identify emerging frontiers and formulate original questions (J. Wang & Shibayama, 2022). Second, thematic mobility, referring to the ability to maintain a coherent intellectual core while flexibly navigating across research contexts and integrating diverse knowledge domains (Y. Wang et al., 2026). Third, network agency, namely the competence to proactively build, maintain, and expand heterogeneous scholarly ties beyond the boundaries of one's doctoral lineage (Baker & Lattuca, 2010; Dunn, 2019; Horta & Santos, 2016). Unlike resources, which are often immediately observable, capability formation tends to unfold subtly through everyday practice. Austin (2002) and Golde (2003) have noted that a moderate degree of structural looseness and supervisory space may provide fertile ground for exploration, enabling trainees to transition from merely doing research to genuinely becoming scholars. This insight implies that the effectiveness of mentorship should not be evaluated solely by mentees' outputs at graduation, but also by their ability to sustain independent trajectories amid evolving academic landscapes.

*Peer interactions and laboratory structure*

Recent research suggests that the effects of mentor guidance extend beyond direct interactions with advisors and are embedded in broader laboratory environments and social structures. Compared with faculty mentoring activities alone, interactions with postdoctoral researchers and senior peers within the lab may play a more substantial role in shaping doctoral students' skill development (Feldon et al., 2019). Meanwhile, Ynalvez et al. (2017) found that an intellectually stimulating laboratory social environment was more strongly associated with doctoral students' publication productivity. From this perspective, mentorship operates as a socially distributed process, in which learning, collaboration, and professional development are embedded in the broader research context rather than driven solely by direct advisor guidance(Ruth et al., 2026).

While peer relationships and group structures may play a positive role in mentees' development, Xing et al. (2025) suggest that mentorship resources are inherently constrained within laboratory settings: As advisors typically supervise multiple mentees simultaneously, the time and attention available to each individual are limited, introducing competitive dynamics among trainees who must vie for access to guidance, collaboration, and mentorship opportunities (Luckhaupt et al., 2005). Therefore, mentees trained in larger groups exhibited stronger academic performance if they remained in academia, but faced a lower probability of long-term academic survival.

Building on these insights, mentorship should be understood as a process embedded in laboratory environments where peer interactions simultaneously shape resource access and capability development. In particular, peer group structures introduce both opportunities for resource acquisition, dynamics of competition and co-development among mentees. Motivated by this perspective, this study examines how mentees' experiences within mentorship considering both direct supervision and peer group contexts affect their citation-elite journal publication.

## METHODOLOGY

We utilize a comprehensive dataset of academic mentorship from 1800 to 2020 in the biosciences[1]. The dataset covers 743,176 mentor–mentee relationships. To capture detailed scholarly activities, these records were linked to publication-level information using unique author and work identifiers provided by the OpenAlex database[2]. To ensure the computability of subsequent collaboration-based measures, we further refined the sample by excluding mentor–mentee pairs without at least one coauthored publication prior to the mentee's graduation. The OpenAlex linkage yielded 556,256 mentor–mentee pairs, of which 239,931 had at least one mentor-linked publication record required for the collaboration-based measures. Detailed sample construction and exclusion criteria are reported in Table S1.

Previous studies have documented a lag between the completion and publication of research (Karakose et al., 2016; Spiegel & Toivanen, 2022). We therefore extend the mentorship period through the third year after graduation to include doctoral research published after degree completion. As a robustness check, we vary this extension from one to three years and obtain qualitatively similar results. The post-graduation period begins after effective graduation and extends to the researcher's most recent publication year, with outcomes assessed over 5-, 10-, and 20-year windows and the full span to capture short-, medium- and long-term, and overall career dynamics. In addition to these conventional windows, the landmark design distinguishes a measurement period covering years 1–3 after effective graduation from two later outcome periods, covering years 4–10 and years 6–20. Both temporal schemes are illustrated in Figure 1.

[1] The dataset is available at https://zenodo.org/records/4917086

[2] OpenAlex database: https://openalex.org. As of May 30, 2025, OpenAlex database includes more than 200 million scholarly publications and related metadata.

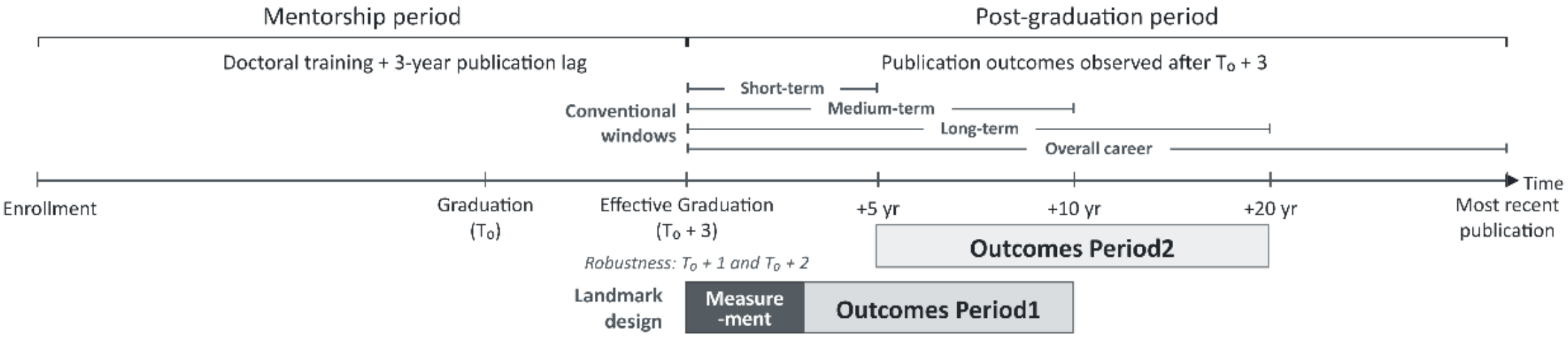

**Figure 1 Temporal definition of mentorship and post-graduation observation windows.**

Based on these temporal definitions, the analysis proceeds in three stages. First, descriptive statistics, Spearman correlations, and trajectory-group comparisons characterize the sample and the distribution of mentor, mentee, and supervision-related variables. Second, propensity-score matching and multivariable regression assess whether these relationships persist after adjustment for measured mentor characteristics, prior mentee performance, and supervision environments. These are complementary observational analyses rather than causal identification strategies. Third, a non-overlapping landmark design separates early post-graduation continuity measures from later publication outcomes.

*Variables*

<u>Outcome variable</u>

To examine one specific dimension of post-training publishing, we focus on the annual rate of publications in journals classified as citation-elite. This outcome captures placement in sources meeting our aggregate citation criteria; it is not a direct measure of article quality, research independence, career retention, or overall scholarly contribution. Using source-level metrics reported in the 2025 OpenAlex snapshot, we define a fixed source set whose two-year mean citedness, source h-index, and total cited-by count each exceed the global 95th-percentile threshold (1.6667, 52.0, and 16,239.6, respectively). The rule identifies 6,847 journals (2.63%; Table S2). Because these source indicators are not field- or publication-year normalized, results should be interpreted as publication in sources with high aggregate citation indicators under this classification rule.

We define **AvgTopPubPerYear_Post** as the average annual number of citation-elite journal publications during the post-graduation period, avoiding the direct dependence of cumulative counts on observed career length.

To characterize heterogeneity more systematically in graduates' citation-elite journal publication trajectories during the supervision period and after that period, we classify graduates into four mutually exclusive groups based on changes in their average annual citation-elite journal publications:

**Table 1 Group definitions based on annual citation-elite journal publication rate before and after graduation.**

| Group | Criterion | Proportion | Description |
|---|---|---|---|
| G1a | m > 0 → n > m | 47.9% | Graduates who already published in citation-elite journals during the supervision period and achieved more annual citation-elite journal publication rate after graduation. |
| G1b | m = 0 → n > 0 | 12.9% | Graduates with no citation-elite journal publications during the supervision period who succeeded in publishing in citation-elite journals after graduation. |
| G2a | m > 0 → n ≤ m | 27.4% | Graduates who published in citation-elite journals during the supervision period but failed to maintain or improve their publication level after graduation. |
| G2b | m = 0 → n = 0 | 11.8% | Graduates with no citation-elite journal publications in either period. |

Note: m means AvgTopPubPerYear_Mentored while n means AvgTopPubPerYear_Post.

Explanatory variables

Prior research suggests that mentees gain mentorship-related resources mainly through direct collaboration with their mentors and through interactions within the local peer environment. Accordingly, we use **Direct_Percent** and **PeerCount_log** to characterize these two training contexts, respectively, and **Topic_Continuity** and **Net_Consistency** to examine how the intellectual and relational foundations formed during mentorship are retained or reconfigured after graduation.

**Mentor collaboration concentration**. Direct collaboration constitutes a primary setting through which mentees gain access to mentors' expertise, feedback, reputation, publication experience, and professional networks (Committee on Effective Mentoring in STEMM et al., 2019; Rosenfeld & Maksimov, 2022). We measure the intensity of direct mentor–mentee collaboration by the proportion of a mentee's publications during the mentorship period that are coauthored with their mentor:

$$Direct_Percent = \frac{P_{direct}}{P_{total}} \quad (1)$$

where $P_{\mathrm{direct}}$ denotes the number of mentor-mentee coauthored publications during mentorship period, and $P_{\mathrm{total}}$ denotes the number of total mentee publications during mentorship period. This measure captures the extent to which the mentee's research activity is embedded in mentor-mediated collaboration, rather than the absolute amount of support received. Greater embeddedness may accelerate learning, but it may also indicate stronger reliance on mentor-led collaborations rather than independent research (Lindahl et al., 2021; Xing et al., 2024).

**Peers in mentorship**. We account for the complex influence of the mentorship environment on mentee development (L. T. Eby et al., 2008), particularly the peer group size, a factor highlighted by prior work noting its nuanced effects: "Academic mentees thrive in big groups, but survive in small groups (Xing et al., 2025)." We define a focal mentee's peers as other mentees supervised by the same mentor whose mentorship periods overlap in calendar time with that of the focal mentee. Accordingly, PeerCount is the number of such overlapping co-mentees, and the variable used in the analyses is log-transformed:

$$PeerCount_log = log(1 + PeerCount) \quad (2)$$

this variable captures the size of the peer group, which prior research suggests may generate both collaborative benefits and competitive pressures.

**Research topic retention.** We construct each researcher's topic set from OpenAlex topics assigned to their publications and calculate the proportion of mentorship-period topics observed again in the post-graduation period:

$$Topic_Cotinuity = \frac{|T_{mentor} \cap T_{post}|}{|T_{mentor}|} \quad (3)$$

where $T_{mentor}$ and $T_{post}$ denote the sets of OpenAlex topics associated with publications in the mentorship period and the post-graduation period, respectively. Higher values indicate that a larger share of mentorship-period topics reappears after graduation.

**Coauthorship network retention.** We examine whether coauthorship ties observed on mentor–mentee joint publications during mentorship reappear after graduation:

$$Net_Consistency = \frac{|N_{mentor} \cap N_{post}|}{|N_{mentor}|} \quad (4)$$

where $N_{\text{mentor}}$ denotes the set of other coauthors on mentor–mentee joint publications during the mentorship period, and $N_{\text{post}}$ represents those collaborators who continue to coauthor with the mentee after graduation. This measure is the retained share of a selected mentor-linked coauthor set; it does not capture the mentee's full network-building capability. Table 2 summarizes these variables, several of which are mechanically related to publication volume.

**Table 2 Variable Definition.**

| Concept | Variable | Definition |
|---|---|---|
| Research topic continuity | Topic_Continuity | Proportion of mentee's research topics during and after mentorship. |
| Coauthorship network continuity | Net_Consistency | Share of coauthors on mentor–mentee joint publications who later coauthor again with the mentee. |
| Mentor collaboration dependence | Direct_Percent | Proportion of mentee's publications during mentorship coauthored with mentor. |
| Peers in mentorship | PeerCount_log | Log-transformed count of peers within the mentorship environment during the mentoring period. |

Control variables

To adjust for measured differences and improve the precision and interpretability of the estimates, we control for **mentees' total publications** and **average annual citation-elite journal publications during the mentorship period**. Mentor ability is included as a control, measured primarily by the **mentor's h-index** at the start of the mentorship. Finally, since mentees with more publications tend to accumulate more coauthors, and the size of the initial network may also be related to the proportion of collaborators retained after graduation, we control for **the size of the mentee's coauthorship network during the mentorship period**.

*Propensity score matching*

Propensity score matching (PSM) is used to improve the comparability of mentor–mentee pairs with different publication trajectories by balancing observed covariates between comparison groups (Ho et al., 2007; Rosenbaum & Rubin, 1983). The unit of analysis is one unique mentor–mentee relationship, denoted by $i$. The propensity score is defined as:

$$e(X_i) = Pr(T_i = 1 \mid X_i) \quad (5)$$

where $T_i$ denotes treatment assignment and $X_i$ includes mentor and mentorship environment characteristics measured prior to post-graduation outcomes.

PSM is used here as a preprocessing procedure to improve balance on observed covariates between comparison groups. It does not eliminate unmeasured confounding or, by itself, establish causal effects. We implement three complementary matching designs for distinct descriptive objectives.

First, a binary-treatment PSM (PSM0) examines whether pre-graduation citation-elite journal publication is associated with differential access to mentoring resources during the supervision period. Treatment is defined as:

$$T_i^{(0)} = I\left(\text{AvgTopPubPerYear}_{i,mentorship} > 0\right) \quad (6)$$

where $\text{AvgTopPubPerYear}_{i,mentorship}$ denotes the annual rate of citation-elite journal publications prior to graduation. Matching is used to assess whether early publication success is associated with differences in mentoring-period characteristics, specifically direct mentee–mentor coauthorship intensity and peer group size.

Second, a binary-treatment PSM (PSM1) evaluates whether the ability to publish in citation-elite journals after graduation represents a structural breakpoint in academic trajectories. Treatment status is defined as:

$$T_i^{(1)} = I\left(\text{AvgTopPubPerYear}_{i,post} > 0\right) \tag{7}$$

Matching is applied to examine whether this threshold outcome is systematically associated with differences in explanatory variables after accounting for observable confounders.

Third, an intensity-based PSM (PSM2) focuses on heterogeneity in post-graduation research output among graduates who successfully publish after graduation. In this design, treatment is defined based on whether a focal explanatory variable exceeds its median value:

$$T_i^{(2)} = I\left(X_{ik} > median(X_k)\right) \tag{8}$$

where $X_{ik}$ denotes the value of explanatory variable $k$ for individual $i$. The outcome of interest is post-graduation publication intensity, measured as $\text{AvgTopPubPerYear}_{i,post}$.

The matched estimand is reported using the conventional average-treatment-on-the-treated (ATT) notation:

$$ATT = E\left[\, Y_i(1) - Y_i(0) \mid T_i^{(2)} = 1 \right] \tag{9}$$

where $Y_i(1)$ and $Y_i(0)$ denote potential outcomes under treatment and control conditions, respectively.

Propensity scores are estimated using logistic regression. Both linear and squared terms of the matching covariates are included in the propensity model. Observations outside common support are removed. Treated observations are matched one-to-one to controls using standardized Euclidean covariate distance within a propensity-score logit caliper. Matching is conducted with replacement, and a treated pair cannot be matched to a control pair belonging to the same mentee. Statistical inference for matched mean differences uses standard errors clustered at the mentee level.

All three matching designs are interpreted as adjusted observational comparisons. PSM2 reports matched mean differences between observations above and below the median of each focal variable among observations with positive post-graduation output. Because the focal variables are not externally assigned and the analysis conditions on observed post-graduation publication, PSM2 does not support a causal ATT interpretation. Detailed post-matching balance diagnostics for the 3-year publication-lag specification are reported in Table S5–Table S7 (see Table S8–Table S10 for robustness checks across lag specifications).

*Regression*

The matching and regression analyses are implemented as parallel and complementary analyses. Regression models are estimated using the complete-case pair-level analytical samples rather than the PSM-matched samples. Because post-graduation citation-elite journal publication counts contain zero values and are strongly right-skewed, the principal specification is a Poisson pseudo-maximum-likelihood model (PPML). For mentor–mentee pair $i$ and outcome window $h$, the model is specified as:

$$\begin{aligned}\log E\left(Y_{i,h} \mid Z_{i,h}, X_i\right) &= \log\left(\text{Exposure}_{i,h}\right) + \beta_1 \text{Topic_Continuity}_{i,h} \\ &+ \beta_2 \text{Net_Consistency}_{i,h} + \beta_3 \text{Direct_Percent}_{i,h} \\ &+ \beta_4 \text{PeerCount_log}_{i,h} + X_i^{'} \gamma + \delta_{g(i)} \qquad (10)\end{aligned}$$

where $Y_{i,h}$ denotes the number of citation-elite journal publications and $Exposure_{i,h}$ denotes the number of observed years in the outcome window. The logarithm of exposure is included as an offset. The vector $Z_{i,h}$ and $X_i$ denote the vectors of explanatory variables and control variables, respectively. The term $\delta_{g(i)}$ represents effective-graduation-year fixed effects. Individual fixed effects are not included because several focal variables do not vary within a pair across observations and would therefore be absorbed. Standard errors in the principal PPML specification are clustered at the mentee level.

As a benchmark, we also estimate OLS models using the annualized citation-elite journal publication rate:

$$\begin{aligned} \mathrm{AvgTopPubPerYear}_{i,h} &= \beta_1\, \mathrm{Topic_Continuity}_{i,h} + \beta_2\, \mathrm{Net_{Consistecny}}_{i,h} \\ &+ \beta_3\, \mathrm{Direct_{Percent}}_{i,h} + \beta_4\, \mathrm{PeerCount_log}_{i,h} + X_i^{'}\gamma + \delta_{g(i)} \\ &+ \varepsilon_{i,h} \end{aligned} \tag{11}$$

For OLS, we report standard errors clustered at the mentee level and, as a sensitivity analysis, standard errors two-way clustered at the mentee and mentor levels. Both explanatory and control variables are entered simultaneously. Variance inflation factor (VIF) diagnostics indicate no problematic multicollinearity among the included predictors (Table S11). Full OLS and PPML estimates are provided in Table S12. Pair-level sample sizes and model diagnostics are reported in Table S17.

*Two-part model*

As an additional robustness analysis, we use a two-part model for outcomes with a mass at zero and a skewed positive distribution to distinguish between the probability of achieving any citation-elite journal publication and publication intensity among positive outcomes (Duan et al., 1983). The first part estimates a linear probability model:

$$D_{i,h} = I\left(Y_{i,h} > 0\right) = Z_{i,h}^{'}\beta + X_i^{'}\gamma + \delta_{g(i)} + \varepsilon_{i,h} \tag{12}$$

The second part is estimated among observations with positive citation-elite journal publication rate using a Gamma generalized linear model and a log link:

$$\log E\left(\mathrm{AvgTopPubPerYear}_{i,h} \mid Y_{i,h} > 0\right) = Z_{i,h}^{'}\beta + X_i^{'}\gamma + \delta_{g(i)} \tag{13}$$

Standard errors in both parts are clustered at the mentee level. Full two-part model results are reported in Table S13. VIF diagnostics are also reported in Table S11, indicating no problematic multicollinearity.

*Landmark design*

To establish a clearer temporal order between continuity measures and subsequent publication outcomes, we additionally implement a non-overlapping landmark design. As illustrated in Figure 1, Topic_Continuity and Net_Consistency are measured during the landmark measurement period, while citation-elite journal publication outcomes are subsequently observed over two outcome periods.

The PPML, two-part, and PSM specifications are re-estimated using these landmark samples. This design separates the measurement of continuity measures from that of subsequent outcomes, although the estimates remain observational. Detailed landmark PPML and PSM2 results are reported in Table S14. Landmark two-part results are presented in Table S15, and the corresponding PSM0 and PSM1 matched comparisons are reported in Table S16.

# RESULTS

*Citation-elite journal publication across two periods*

Table 3 presents the Spearman correlations between explanatory, control variables and the outcome variable across 5-, 10-, and 20-year outcome windows. Most explanatory variables exhibit relatively weak monotonic associations with post-graduation outcomes, with absolute correlation coefficients generally below 0.2. In contrast, pre-graduation performance metrics—specifically, the annual citation-elite journal publication rate (AvgTopPubPerYear_Mentored) and the annual overall publication rate (AvgPaperPerYear_Mentored) during the mentoring period—show substantially stronger correlations with annual post-graduation citation-elite journal publication rate (AvgTopPerYear_Post). This may indicate a notable persistence in academic output across the graduation boundary, whereby mentees who publish more prior to graduation also tend to exhibit higher productivity afterward. However, such aggregate correlations alone cannot reveal whether this persistence reflects continued access to mentor-related resources, cumulative advantage, or the internalization of research capability; therefore, we further examine these relationships through group-based comparisons and adjusted observational analyses.

**Table 3 Spearman Correlation with AvgTopPubYear_Post.**

| Variable | Post-graduation outcome window | | |
|---|---|---|---|
| | **5 years** | **10 years** | **20 years** |
| Topic_Continuity | +0.230*** | +0.161*** | +0.075*** |
| Net_Consistency | +0.311*** | +0.290*** | +0.263*** |
| Direct_Percent | −0.101*** | −0.072*** | −0.008 |
| PeerCount_log | −0.059*** | −0.060*** | −0.057*** |
| h_index | +0.174*** | +0.168*** | +0.162*** |
| AvgPaperPerYear_Mentored | +0.813*** | +0.818*** | +0.809*** |
| NetworkSize_Mentored | +0.565*** | +0.571*** | +0.569*** |
| AvgTopPubPerYear_Mentored | +0.822*** | +0.818*** | +0.794*** |

As described in Table 1, we classify mentees into four groups based on their citation-elite journal publication trajectories. We then examine explanatory and control variables across these groups to assess how mentor characteristics, mentee attributes, and supervisory environments are differentially associated with pre- and post-graduation citation-elite journal publication trajectories. Figure 2 provides a descriptive comparison of the four groups, showing that mentees who maintain citation-elite journal publication performance across both periods and those who publish only in one period share partially overlapping distributions on certain dimensions, but differ on others. Specifically, differences between G1 and G2 appear in both explanatory and control variables. Mentees in G1 show lower Topic_Continuity and Direct_Percent than those in G2, while Panels B and C show that G1 is associated with mentors with higher h-index values and larger mentored collaboration networks. Net_Consistency displays a more complex pattern: G1 shows higher Net_Consistency than G2 when pre-graduation publication status is held constant, whereas subgroup b has higher Net_Consistency than subgroup a within both G1 and G2. Moreover, compared with subgroup b, subgroup a displays higher Direct_Percent and PeerCount_log, suggesting that mentees with pre-graduation citation-elite journal publications tend to have more intensive direct collaboration with mentors and larger peer group size during the mentorship period. Detailed group-wise descriptive statistics are reported in Table S4.

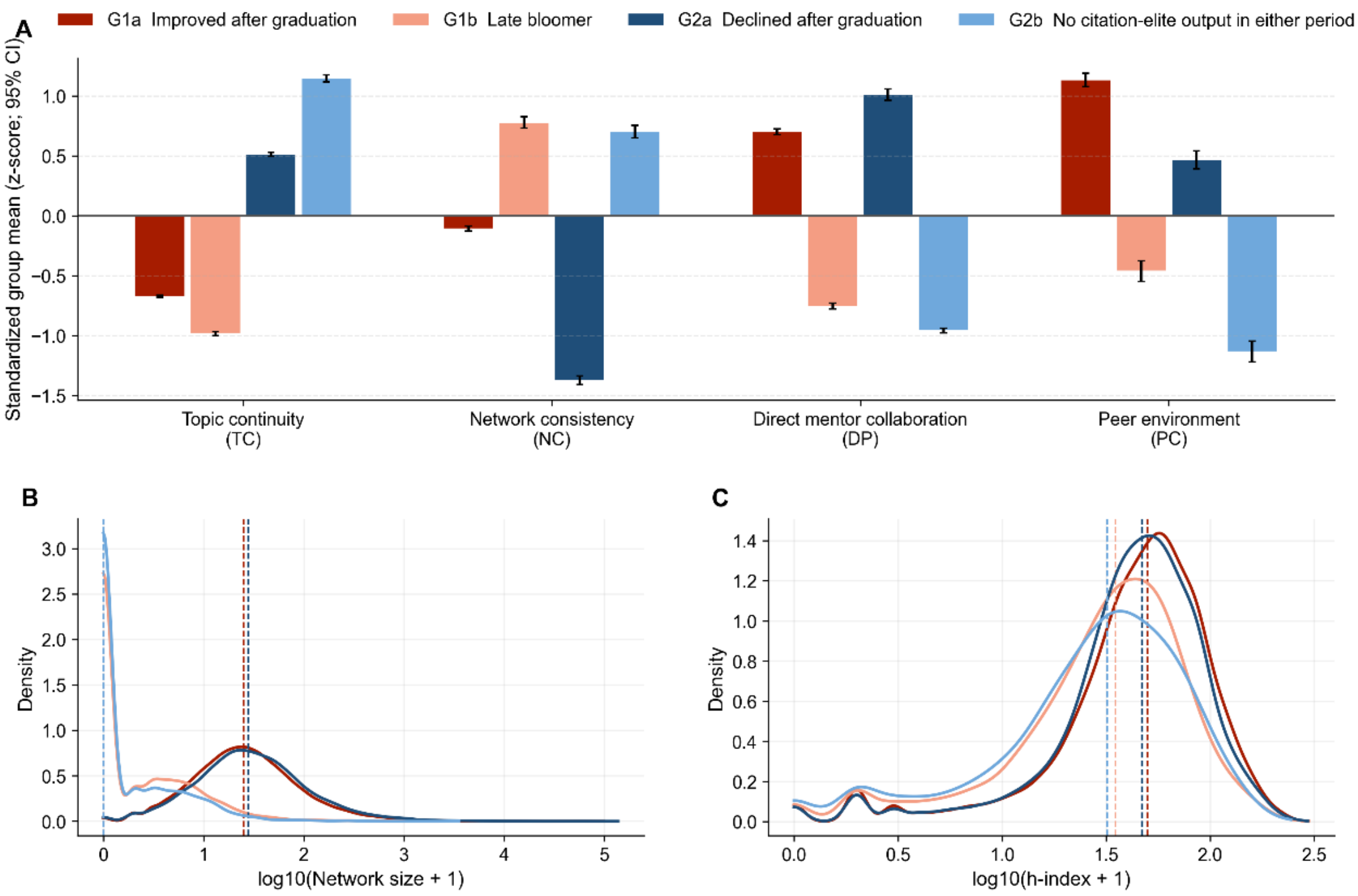

**Figure 2 Distributional comparison of explanatory and control variables across four groups. Note: Panel (A) reports standardized group means with 95% confidence intervals for the four explanatory variables. Panels (B) and (C) display kernel density distributions of mentorship period network size and mentor h-index, respectively. Vertical dashed lines indicate group-specific medians.**

Overall, the descriptive results show a positive association between pre-graduation and post-graduation citation-elite journal publication performance, indicating some persistence across career stages, while the association is incomplete: Mentees with pre-graduation citation-elite journal publications, whether they sustain this performance after graduation, consistently show higher Direct_Percent and PeerCount_log. At the same time, some mentees without pre-graduation citation-elite journal publications still publish in citation-elite journals after graduation, and these mentees resemble sustained performers on Topic_Continuity and Net_Consistency.

Moreover, the observed patterns for collaboration-related variables, such as network consistency, indicate that their apparent associations may be confounded by differences in collaboration scale. These considerations motivate a more integrated analytical approach; accordingly, following sections employ regression-based analyses and propensity score matching to jointly evaluate these factors and to distinguish the roles of resource acquisition and capability internalization more clearly in shaping post-graduation citation-elite journal publication performance.

*Resource-related factors and citation-elite journal publication*

The PSM0 results in Figure 3 show that, after matching on mentor characteristics, mentorship-period productivity, and collaboration scale, mentees with mentorship-period citation-elite journal publications show higher Direct_Percent and PeerCount_log (SMD = +0.47 and +0.13). The results are consistent with the unmatched comparisons presented in Figure 2, suggesting that stronger direct mentor involvement and a larger peer group size characterize mentees who achieve citation-elite journal publication during training. Such conditions may provide greater access to guidance, collaboration opportunities, and other mentorship-related resources, consistent with a Matthew-effect–type pattern in which structural affiliation may contribute to early publication advantage (Merton, 1968). Detailed window-specific PSM0 comparisons and balance diagnostics are reported in Table S5, with publication-lag robustness results reported in Table S8.

In the overall PSM1 comparison, post-graduation publishers have lower Direct_Percent and PeerCount_log during mentorship. The same negative direction appears across the 5-, 10-, and 20-year post-graduation windows (Table S6), and is robust to the 1-, 2-, and 3-year publication-lag definitions (Table S9). The two-part model estimates are generally negative for both variables. For a 0.1-unit increase in Direct_Percent, the estimated probability differences are -0.11, -0.09, and -0.08 percentage points across the 5-, 10-, and 20-year windows; the 20-year estimate is not statistically distinguishable from zero (Table S13). Corresponding intensity changes among positive outcomes are -4.0%, -3.6%, and -2.3%. PeerCount_log estimates are approximately -0.06 percentage points on probability and -0.5%, -0.7%, and -1.0% on intensity. These observational associations do not establish that mentor collaboration or peer-group size causes later constraints.

Thus, mentor-coauthorship concentration and overlapping co-mentee group size characterize different publication trajectories across career stages. They are positively associated with training-period citation-elite publishing and generally negatively associated with post-training outcomes, but neither measure directly captures mentoring support, dependence, or autonomy.

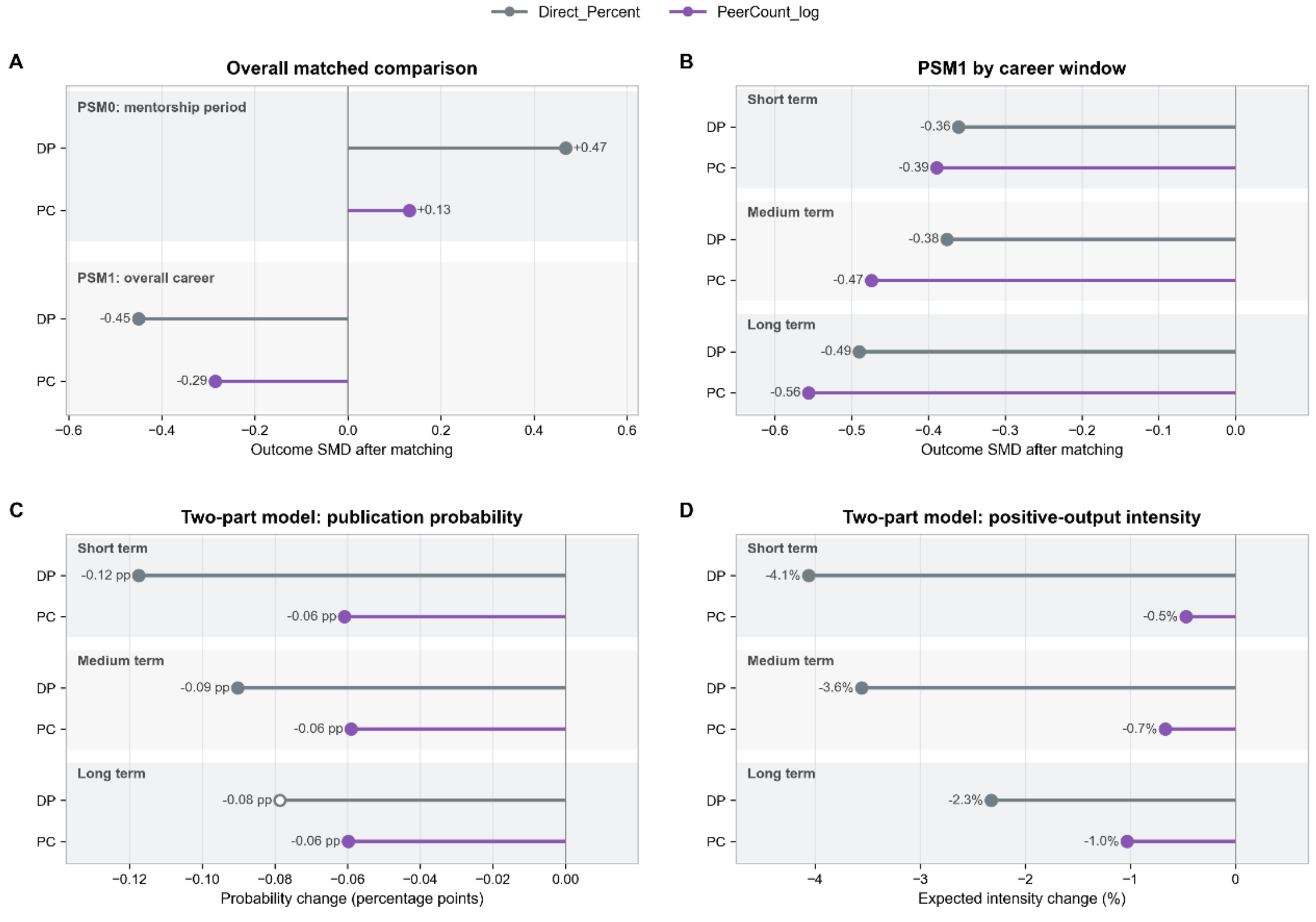

**Figure 3 Differences in resource-related variables between matched groups under PSM0 and PSM1. Note: Panel (A) presents standardized mean differences (SMDs) in Direct_Percent and PeerCount_log for the overall mentorship-period outcome under PSM0 and the overall post-graduation outcome under PSM1. Panel (B) shows PSM1 results across short-, medium-, and long-term post-graduation windows. Panel (C) reports changes in the probability of any post-graduation citation-elite journal publication associated with a 0.1-unit increase in each variable. Panel (D) reports the corresponding changes in publication intensity among observations with positive citation-elite journal publication rate. The PSM estimates represent matched descriptive differences rather than causal effects. The hollow marker indicates $p \geq .05$. DP = Direct_Percent; PC = PeerCount_log.**

*Capability-related factors and citation-elite journal publication*

Figure 4 reports landmark and conventional continuity analyses. In landmark PPML models, early Topic_Continuity and Net_Consistency are positively associated with subsequent publication counts in both outcome windows. Margin-specific results are less uniform: for years 6–20, Topic_Continuity is not statistically distinguishable from zero in either part of the two-part model, and the PSM1 difference in Net_Consistency is also nonsignificant. Among observations with positive output, landmark PSM2 matched differences for Topic_Continuity and Net_Consistency are +0.51 and +0.24 in years 4–10 and +0.55 and +0.57 in years 6–20; the corresponding PPML rate changes for a 0.1-unit increase are +3.7%/+5.4% and +2.5%/+2.7%. Conventional PSM2 estimates for Topic_Continuity shift from +0.35 in the short term to -0.29 and -1.03 in the medium and long windows, while Net_Consistency remains positive (+0.30, +0.34, and +0.51). Detailed conventional PSM2 comparisons and balance diagnostics are reported in Table S7, with publication-lag robustness results reported in Table S10.

These measures display distinct temporal patterns. Early topic retention may lower transition costs, but the negative conventional estimates over longer same-period windows do not demonstrate intentional agenda broadening because continuity and output are measured contemporaneously. Retained collaborator share is more consistently associated with publication intensity. Existing ties may facilitate coordination and access to complementary expertise as topics change (Reagans & McEvily, 2003; Tripodi et al., 2020), but the measure captures observed continuity rather than autonomous network agency.

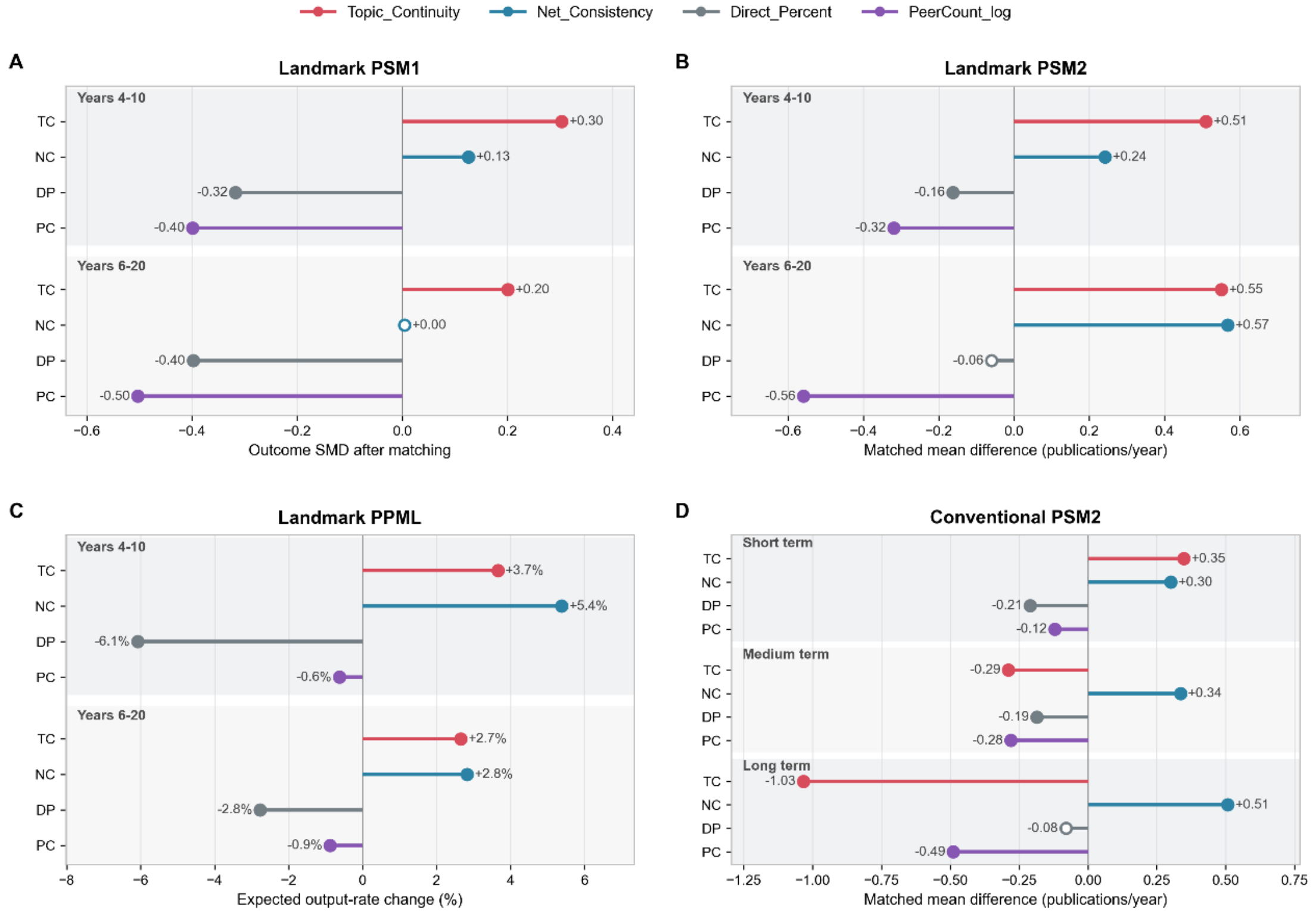

**Figure 4 Associations between explanatory variables and post-graduation publication outcomes under conventional and landmark analyses. Note: Panel (A) presents post-matching differences in explanatory variables between publication-status groups under landmark PSM1. Panel (B) reports matched differences in annual citation-elite journal publication rate under landmark PSM2. Panel (C) presents the corresponding landmark PPML estimates. Panel (D) reports conventional PSM2 results across short-, medium-, and long-term post-graduation windows. Filled markers indicate $p < .05$, whereas hollow markers indicate $p \geq .05$. All matching specifications satisfy max |covariate SMD| < .10. Abbreviated variable labels follow the same convention as in Figure 3: TC = Topic_Continuity, NC = Network_Consistency.**

## DISCUSSION AND CONCLUSION

In this paper, we use propensity score matching and regression-based analyses to examine how several mentorship-related factors are associated with mentees' citation-elite journal publication, including research topic continuity, direct mentor collaboration, coauthorship network continuity and peer group size. We find that citation-elite journal publication during mentorship is strongly associated with subsequent citation-elite journal publication, but the factors associated with success vary across career stages. Direct mentor involvement and larger peer group size are associated with stronger publication performance during training, whereas these relationships weaken or reverse after graduation. By contrast, post-graduation performance is more closely associated with how intellectual and relational foundations accumulated during training are carried forward, although topic and network continuity display different temporal patterns. These findings point to a stage-dependent role of mentorship resources. Resources that facilitate publication during training do not necessarily remain advantageous once formal supervision recedes. Their longer-term value appears to depend partly on whether mentees can develop greater research autonomy and use, adapt, or reconfigure the intellectual and relational foundations accumulated during training in a more independent research context (Patsali et al., 2024; Rosenfeld & Maksimov, 2022).

We begin by examining the relationship between citation-elite journal publishing during mentorship and after graduation. Consistent with prior research (Pinheiro et al., 2014; Horta & Santos, 2016; Acevedo et al., 2024), we find robust evidence of a persistence pattern: mentees who publish in citation-elite journals during mentorship are more likely to sustain citation-elite journal publication post-graduation, but this persistence may reflect selection, cumulative advantage, or processes established during training; the observational design cannot distinguish among these mechanisms. Resource-related factors, such as mentorship support and collaboration opportunities, are more strongly associated with citation-elite journal publication during mentorship, in line with prior studies emphasizing the role of mentorship in fostering initial research productivity (Gutierrez et al., 2021).

However, the patterns associated with success differ across the two periods. Characteristics positively associated with citation-elite journal publication during mentorship do not always translate into higher post-graduation output. Some mentees who did not publish in citation-elite journals during mentorship nevertheless achieve post-graduation citation-elite journal publications, and these individuals resemble successful mentees in terms of other dimensions, particularly in their ability to mobilize established knowledge and collaboration networks. This suggests that while early publication provides a foundation for later success, different dimensions of mentorship and research development may evolve independently across career stages, consistent with prior observations on the complex and multidimensional nature of academic development (Gardner, 2007).

A larger overlapping co-mentee group is positively associated with citation-elite journal publishing during the training window but negatively associated with the same outcome afterward. Because the analysis does not observe employment status or reasons for leaving publication, we interpret this pattern as a difference in observable publishing trajectories rather than academic survival. The association may reflect resource access, competition, coordination, selection into different research settings, or other unmeasured features of laboratory organization.

A similar transformation is observed in direct mentor collaboration. During the mentorship period, higher mentor collaboration concentration is positively associated with citation-elite journal publication, which may provide close mentor involvement and support, as well as the visibility advantages brought by the mentor's academic reputation and network (Committee on Effective Mentoring in STEMM et al., 2019; L. T. Eby et al., 2000; L. T. D. T. Eby et al., 2013). However, while strong mentor involvement may enable short-term publication success, it may also limit the development of independent research capability (Baker & Lattuca, 2010; Pol, 2025; J. Wang & Shibayama, 2022; Y. Wang et al., 2026). This is because resource-driven advantages acquired during training do not automatically translate into long-term productivity, and one possible interpretation is that mentor-linked collaboration becomes less portable after formal supervision ends; however, the study does not directly measure autonomy or independent management of the research process, including problem formulation, execution, and publication (Austin, 2002; Golde, 2003; Ma et al., 2020).

After graduation, mentorship resources appear to take on a different role as mentees move into a more autonomous research context, so we focus on mentees' independent research capability, particularly the ability to reconfigure and deploy previously accumulated resources under greater autonomy. Our findings suggest that maintaining some continuity with earlier research topics can provide a useful foundation for post-graduation publication, particularly during the transition to independent research, while established collaborative ties remain valuable as research agendas continue to evolve. As direct supervisory support recedes, sustained research performance may therefore require a balance between continuity and adjustment. Prior knowledge can reduce the costs of entering an independent career, but longer-term development may also require researchers to move beyond established topics and pursue new directions. Collaborative relationships, by contrast, can remain useful across changing research agendas because accumulated trust, coordination experience, and complementary expertise can be redeployed across different intellectual trajectories (Reagans & McEvily, 2003; Tripodi et al., 2020). Meanwhile, these findings further suggest that early disadvantages associated with resource-dependent training are not necessarily persistent. Through effective utilization and extension of previously accumulated resources in the post-graduation stage, mentees may still be able to exhibit post-graduation publication trajectories that are not fully determined by mentorship-period publication status (Brownrout et al., 2021; Nicolaisen, 2025).

This study contributes a longitudinal description of how mentor-coauthorship concentration, overlapping co-mentee group size, topic retention, and retained-collaborator share are associated with citation-elite journal publishing across post-training windows. For information science, it illustrates how scholarly-communication traces can illuminate resource conversion across academic careers while also exposing the limits of journal-level indicators as measures of individual quality. The results do not directly identify resource transfer, latent capability, or causal developmental pathways, and they should not be used to recommend less mentor collaboration or smaller groups. A more defensible practical implication is to pair resource-rich training with mechanisms that make support portable after training, including multiple mentors, transparent authorship and credit, opportunities for agenda ownership, broader external networks, and transition support.

This study has several limitations. First, all estimates are observational; PSM,

regression, and landmark designs address measured imbalance or temporal ordering but do not establish causation. Second, analyses condition on successful bibliographic linkage, mentor-linked publication activity, window eligibility, and complete data. The Academic Family Tree is crowdsourced and contains multiple mentorship types, while longer windows disproportionately represent earlier cohorts and researchers who remain observable in publication records. Third, coauthorship concentration, overlapping co-mentee counts, topic retention, and retained-collaborator share are proxies rather than direct measures of support, peer interaction, autonomy, or capability. Fourth, the fixed citation-elite classification uses globally pooled source indicators and may reflect field, journal-age, and snapshot-time differences; venue placement is not article-level quality. Finally, the data do not identify career retention, employment sector, or whether associations differ by gender, geography, institutional prestige, citizenship, caregiving, or other structural conditions. Future work should test field- and year-normalized outcomes, cohort heterogeneity, and qualitative mechanisms.

## ACKNOWLEDGMENTS

This work was supported by the National Natural Science Foundation of China (#72474009 and # L252400109).

## GENERATIVE AI STATEMENTS

During the preparation of this manuscript, the authors used ChatGPT, developed by OpenAI, to improve language, clarity, and readability. All AI-assisted text was subsequently reviewed and revised by the authors, who take full responsibility for the accuracy and final content of the manuscript.

## CODE AND DATA AVAILABILITY

Code can be found at: https://github.com/zedthebolter/Mentorship_Resources. Data are all open to public.

# SUPPLEMENTARY INFORMATION

**Table S1. Sample construction and exclusion criteria.**

| Step | Sample restriction | N remaining | N excluded |
|---|---|---|---|
| Original mentorship records | Bioscience mentor–mentee relationships, 1800–2020 | 743,176 | — |
| Linked to OpenAlex with usable publication records | Successfully matched to OpenAlex author/work identifiers and retained publication-level metadata required for analysis | 556,256 | 186,920 |
| Has at least one mentor-linked publication record required for collaboration-based measures | Applied to retain records usable for collaboration-based variable construction | 239,931 | 316,325 |

**Table S2. Identification of citation-elite journals in OpenAlex.**

Panel A. Thresholds for individual indicators

| Indicator | Threshold (95th percentile) | Journals meeting threshold | Share of all journals |
|---|---|---|---|
| 2yr_mean_citedness | 1.6667 | 13,051 | 5.00% |
| h_index | 52.0000 | 13,195 | 5.06% |
| cited_by_count | 16,239.6000 | 13,040 | 5.00% |

Panel B. Combined classification rule

| Item | Value |
|---|---|
| Total number of journals in OpenAlex | 260,798 |
| Journals meeting none of the three thresholds | 240,425 |
| Journals meeting one threshold | 8,307 |
| Journals meeting two thresholds | 5,219 |
| Journals meeting all three thresholds | 6,847 |
| Final citation-elite journal set | 6,847 |
| Share of all journals | 2.63% |

**Table S3. Descriptive statistics of all variables across publication-lag specifications.**

Panel A. 1-year publication lag

| Variable | 5-year window | | | 10-year window | | | 20-year window | | | Overall Career | | |
|---|---|---|---|---|---|---|---|---|---|---|---|---|
| N | 58,501 | | | 46,394 | | | 22,234 | | | 60,969 | | |
| | Mean | Median | Std | Mean | Median | Std | Mean | Median | Std | Mean | Median | Std |
| Topic_Continuity | 0.175 | 0.154 | 0.151 | 0.138 | 0.119 | 0.120 | 0.092 | 0.075 | 0.088 | 0.117 | 0.091 | 0.114 |
| Network_Consistency | 0.164 | 0.000 | 0.260 | 0.170 | 0.000 | 0.272 | 0.153 | 0.000 | 0.275 | 0.181 | 0.000 | 0.272 |
| Direct_Percent | 0.402 | 0.381 | 0.340 | 0.391 | 0.353 | 0.345 | 0.354 | 0.286 | 0.347 | 0.404 | 0.385 | 0.342 |
| PeerCount_log | 0.955 | 0.693 | 0.955 | 0.955 | 0.693 | 0.960 | 0.771 | 0.693 | 0.886 | 0.955 | 0.693 | 0.955 |
| AvgTopPubPerYear_Mentored | 1.318 | 0.833 | 2.093 | 1.202 | 0.750 | 1.667 | 1.067 | 0.667 | 1.405 | 1.320 | 0.833 | 2.126 |
| AvgPaperPerYear_Mentored | 2.766 | 1.714 | 4.649 | 2.496 | 1.500 | 3.936 | 2.104 | 1.333 | 2.514 | 2.786 | 1.667 | 5.607 |
| h_index | 54.026 | 48.000 | 36.377 | 55.346 | 49.000 | 36.935 | 55.685 | 50.000 | 37.750 | 53.726 | 47.000 | 36.265 |
| NetworkSize_Mentored | 68.083 | 16.000 | 958.438 | 48.533 | 13.000 | 481.833 | 25.734 | 9.000 | 125.468 | 68.576 | 16.000 | 943.007 |
| AvgTopPubPerYear_Post | 2.057 | 1.400 | 3.289 | 2.293 | 1.500 | 3.300 | 2.720 | 1.800 | 3.267 | 2.605 | 1.576 | 4.006 |

Panel B. 2-year publication lag

| Variable | 5-year window | | | 10-year window | | | 20-year window | | | Overall career | | |
|---|---|---|---|---|---|---|---|---|---|---|---|---|
| N | 57,751 | | | 44,736 | | | 21,440 | | | 61,981 | | |
| | Mean | Median | Std | Mean | Median | Std | Mean | Median | Std | Mean | Median | Std |
| Topic_Continuity | 0.175 | 0.154 | 0.144 | 0.142 | 0.125 | 0.116 | 0.099 | 0.083 | 0.087 | 0.122 | 0.100 | 0.111 |
| Network_Consistency | 0.136 | 0.000 | 0.235 | 0.143 | 0.000 | 0.248 | 0.129 | 0.000 | 0.251 | 0.153 | 0.000 | 0.248 |
| Direct_Percent | 0.373 | 0.333 | 0.320 | 0.363 | 0.333 | 0.324 | 0.328 | 0.250 | 0.323 | 0.377 | 0.339 | 0.321 |
| PeerCount_log | 1.019 | 0.693 | 0.989 | 1.011 | 0.693 | 0.994 | 0.818 | 0.693 | 0.923 | 1.015 | 0.693 | 0.989 |
| AvgTopPubPerYear_Mentored | 1.348 | 0.857 | 2.132 | 1.217 | 0.800 | 1.660 | 1.088 | 0.714 | 1.419 | 1.359 | 0.857 | 2.177 |
| AvgPaperPerYear_Mentored | 2.819 | 1.750 | 4.972 | 2.517 | 1.571 | 3.912 | 2.149 | 1.423 | 2.518 | 2.873 | 1.750 | 5.710 |
| h_index | 54.157 | 48.000 | 36.482 | 55.350 | 49.000 | 37.096 | 55.126 | 49.000 | 37.675 | 53.619 | 47.000 | 36.308 |
| NetworkSize_Mentored | 80.636 | 20.000 | 1078.612 | 56.360 | 16.000 | 542.745 | 30.110 | 11.000 | 140.888 | 84.400 | 20.000 | 1103.041 |
| AvgTopPubPerYear_Post | 2.188 | 1.400 | 3.518 | 2.400 | 1.500 | 3.534 | 2.776 | 1.800 | 3.397 | 2.642 | 1.591 | 4.090 |

Panel C. 3-year publication lag

| **Variable** | **5-year window** | | | **10-year window** | | | **20-year window** | | | **Overall career** | | |
|---|---|---|---|---|---|---|---|---|---|---|---|---|
| N | | 56,070 | | | 42,823 | | | 20,568 | | | 62,394 | |
| | Mean | Median | Std | Mean | Median | Std | Mean | Median | Std | Mean | Median | Std |
| Topic_Continuity | 0.176 | 0.162 | 0.138 | 0.148 | 0.133 | 0.114 | 0.106 | 0.091 | 0.088 | 0.127 | 0.111 | 0.109 |
| Network_Consistency | 0.115 | 0.000 | 0.214 | 0.124 | 0.000 | 0.229 | 0.114 | 0.000 | 0.233 | 0.131 | 0.000 | 0.228 |
| Direct_Percent | 0.338 | 0.300 | 0.299 | 0.327 | 0.278 | 0.302 | 0.293 | 0.222 | 0.298 | 0.344 | 0.308 | 0.301 |
| PeerCount_log | 1.073 | 1.099 | 1.020 | 1.058 | 0.693 | 1.022 | 0.857 | 0.693 | 0.948 | 1.066 | 0.693 | 1.017 |
| AvgTopPubPerYear_Mentored | 1.386 | 0.889 | 2.158 | 1.259 | 0.833 | 1.724 | 1.122 | 0.750 | 1.467 | 1.419 | 0.900 | 2.260 |
| AvgPaperPerYear_Mentored | 2.914 | 1.800 | 5.337 | 2.601 | 1.625 | 4.174 | 2.217 | 1.500 | 2.547 | 3.015 | 1.833 | 5.882 |
| h_index | 54.374 | 48.000 | 36.661 | 55.417 | 49.000 | 37.325 | 54.699 | 49.000 | 37.542 | 53.604 | 47.000 | 36.422 |
| NetworkSize_Mentored | 93.928 | 24.000 | 1178.501 | 65.406 | 19.000 | 624.557 | 35.641 | 13.000 | 160.911 | 102.818 | 25.000 | 1258.936 |
| AvgTopPubPerYear_Post | 2.308 | 1.400 | 3.734 | 2.509 | 1.600 | 3.752 | 2.839 | 1.850 | 3.546 | 2.681 | 1.600 | 4.188 |

**Table S4. Group-wise descriptive statistics for G1a, G1b, G2a, and G2b.**

| Variable | G1a | | G1b | | G2a | | G2b | |
|---|---|---|---|---|---|---|---|---|
| N | 112,394 | | 15,623 | | 95,732 | | 16,182 | |
| | Mean | Median | Mean | Median | Mean | Median | Mean | Median |
| AvgTopPubPerYear_Mentored | 0.822 | 0.500 | 0.000 | 0.000 | 0.986 | 0.667 | 0.000 | 0.000 |
| AvgTopPubPerYear_Post | 2.286 | 1.385 | 0.441 | 0.182 | 0.549 | 0.300 | 0.000 | 0.000 |
| Topic_Continuity | 0.214 | 0.184 | 0.171 | 0.125 | 0.331 | 0.297 | 0.317 | 0.250 |
| Network_Consistency | 0.597 | 0.613 | 0.683 | 0.750 | 0.555 | 0.552 | 0.704 | 0.917 |
| PeerCount | 3.725 | 2.000 | 2.085 | 1.000 | 3.422 | 2.000 | 1.877 | 1.000 |
| h_index | 46.676 | 41.000 | 31.075 | 26.000 | 45.155 | 40.000 | 24.053 | 19.000 |
| NetworkSize_Mentored | 108.186 | 25.000 | 11.572 | 5.000 | 73.164 | 17.000 | 9.072 | 3.000 |

**Table S5. PSM0 balance diagnostics across post-graduation windows (t3: 3-year lag).**

| Outcome | Matched pairs | Maximum \|SMD\| after matching | Outcome SMD after matching | p-value (clustered) |
|---|---|---|---|---|
| Direct_Percent | 61,743 | 0.015 | +0.467 | <0.001 |
| PeerCount_log | 61,978 | 0.015 | +0.132 | 0.024 |

**Table S6. PSM1 balance diagnostics across post-graduation windows (t3: 3-year lag).**

| Window | Outcome | Matched pairs | Maximum \|SMD\| after matching | Outcome SMD (clustered p-value) |
|---|---|---|---|---|
| 5-year | Topic_Continuity | 24,419 | 0.034 | +0.628 (<0.001) |
| | Network_Consistency | 51,994 | 0.034 | +0.576 (<0.001) |
| | Direct_Percent | 51,994 | 0.034 | -0.361 (<0.001) |
| | PeerCount_log | 53,263 | 0.034 | -0.389 (<0.001) |
| 10-year | Topic_Continuity | 20,207 | 0.039 | +0.441 (<0.001) |
| | Network_Consistency | 39,881 | 0.039 | +0.523 (<0.001) |
| | Direct_Percent | 39,881 | 0.039 | -0.376 (<0.001) |
| | PeerCount_log | 41,476 | 0.039 | -0.474 (<0.001) |
| 20-year | Topic_Continuity | 9,068 | 0.051 | +0.164 (0.036) |
| | Network_Consistency | 18,573 | 0.051 | +0.411 (<0.001) |
| | Direct_Percent | 18,573 | 0.051 | -0.490 (<0.001) |
| | PeerCount_log | 19,968 | 0.051 | -0.556 (<0.001) |
| Overall career | Topic_Continuity | 23,627 | 0.022 | +0.286 (<0.001) |
| | Network_Consistency | 59,632 | 0.022 | +0.665 (<0.001) |
| | Direct_Percent | 59,632 | 0.022 | -0.450 (<0.001) |
| | PeerCount_log | 61,993 | 0.022 | -0.285 (<0.001) |

**Table S7. PSM2 balance diagnostics (t3: 3-year lag).**

| Window | Outcome | Matched pairs | Maximum \|SMD\| after matching | Outcome SMD (clustered p-value) |
|---|---|---|---|---|
| 5-year | Topic_Continuity | 24,312 | 0.012 | +0.348 (<0.001) |

| Window | Outcome | Matched pairs | Maximum \|SMD\| after matching | Outcome SMD (clustered p-value) |
|---|---|---|---|---|
| | Network_Consistency | 20,161 | 0.011 | +0.301 (<0.001) |
| | Direct_Percent | 26,374 | 0.007 | −0.210 (<0.001) |
| | PeerCount_log | 26,947 | 0.008 | −0.121 (0.003) |
| 10-year | Topic_Continuity | 20,607 | 0.014 | −0.289 (<0.001) |
| | Network_Consistency | 14,985 | 0.013 | +0.336 (<0.001) |
| | Direct_Percent | 20,136 | 0.009 | −0.186 (<0.001) |
| | PeerCount_log | 20,734 | 0.008 | −0.281 (<0.001) |
| 20-year | Topic_Continuity | 9,906 | 0.017 | −1.033 (<0.001) |
| | Network_Consistency | 5,880 | 0.016 | +0.507 (<0.001) |
| | Direct_Percent | 9,962 | 0.018 | −0.080 (0.221) |
| | PeerCount_log | 8,412 | 0.009 | −0.490 (<0.001) |
| Overall career | Topic_Continuity | 30,292 | 0.013 | −0.702 (<0.001) |
| | Network_Consistency | 24,603 | 0.011 | +0.301 (<0.001) |
| | Direct_Percent | 29,728 | 0.007 | −0.329 (<0.001) |
| | PeerCount_log | 30,427 | 0.007 | −0.384 (<0.001) |

**Table S8. Publication-lag robustness of PSM0 results across t1, t2, and t3.**

| Lag t | Outcome | Matched pairs | Maximum \|SMD\| | Outcome SMD (clustered p-value) |
|---|---|---|---|---|
| 1 | Direct_Percent | 56,584 | 0.016 | +0.416 (<0.001) |
| | PeerCount_log | 56,915 | 0.016 | +0.077 (0.102) |
| 2 | Direct_Percent | 59,505 | 0.014 | +0.438 (<0.001) |
| | PeerCount_log | 59,772 | 0.014 | +0.179 (<0.001) |
| 3 | Direct_Percent | 61,743 | 0.015 | +0.467 (<0.001) |
| | PeerCount_log | 61,978 | 0.015 | +0.132 (0.024) |

**Table S9 Publication-lag robustness of PSM1 matched comparisons.**

| Lag t | Window | Outcome | Matched pairs | Max \|SMD\| | Balance | Treated/high mean | Control/low mean | Outcome SMD | p-value (clustered) |
|---|---|---|---|---|---|---|---|---|---|
| 1 | 5-year | Topic_Continuity | 28,498 | 0.030 | Pass | 0.178 | 0.110 | +0.426 | <0.001 |
| 1 | 5-year | Network_Consistency | 54,774 | 0.030 | Pass | 0.169 | 0.037 | +0.631 | <0.001 |
| 1 | 5-year | Direct_Percent | 54,774 | 0.030 | Pass | 0.407 | 0.467 | -0.174 | <0.001 |
| 1 | 5-year | PeerCount_log | 57,655 | 0.030 | Pass | 0.944 | 1.304 | -0.353 | <0.001 |
| 1 | 10-year | Topic_Continuity | 24,868 | 0.038 | Pass | 0.140 | 0.116 | +0.156 | 0.009 |
| 1 | 10-year | Network_Consistency | 44,336 | 0.038 | Pass | 0.173 | 0.040 | +0.609 | <0.001 |
| 1 | 10-year | Direct_Percent | 44,336 | 0.038 | Pass | 0.394 | 0.481 | -0.245 | <0.001 |
| 1 | 10-year | PeerCount_log | 47,852 | 0.038 | Pass | 0.931 | 1.396 | -0.455 | <0.001 |
| 1 | 20-year | Topic_Continuity | 12,755 | 0.068 | Pass | 0.093 | 0.100 | -0.074 | 0.394 |
| 1 | 20-year | Network_Consistency | 21,240 | 0.068 | Pass | 0.152 | 0.057 | +0.418 | <0.001 |
| 1 | 20-year | Direct_Percent | 21,240 | 0.068 | Pass | 0.357 | 0.448 | -0.250 | <0.001 |

| Lag t | Window | Outcome | Matched pairs | Max \|SMD\| | Balance | Treated/high mean | Control/low mean | Outcome SMD | p-value (clustered) |
|---|---|---|---|---|---|---|---|---|---|
| 1 | 20-year | PeerCount_log | 23,851 | 0.068 | Pass | 0.745 | 1.319 | -0.595 | <0.001 |
| 2 | 5-year | Topic_Continuity | 24,460 | 0.030 | Pass | 0.176 | 0.101 | +0.505 | <0.001 |
| 2 | 5-year | Network_Consistency | 53,838 | 0.030 | Pass | 0.140 | 0.024 | +0.630 | <0.001 |
| 2 | 5-year | Direct_Percent | 53,838 | 0.030 | Pass | 0.377 | 0.471 | -0.281 | <0.001 |
| 2 | 5-year | PeerCount_log | 55,721 | 0.030 | Pass | 1.010 | 1.413 | -0.381 | <0.001 |
| 2 | 10-year | Topic_Continuity | 22,026 | 0.029 | Pass | 0.145 | 0.105 | +0.287 | <0.001 |
| 2 | 10-year | Network_Consistency | 41,958 | 0.029 | Pass | 0.145 | 0.030 | +0.591 | <0.001 |
| 2 | 10-year | Direct_Percent | 41,958 | 0.029 | Pass | 0.367 | 0.474 | -0.314 | <0.001 |
| 2 | 10-year | PeerCount_log | 44,293 | 0.029 | Pass | 0.992 | 1.519 | -0.500 | <0.001 |
| 2 | 20-year | Topic_Continuity | 11,241 | 0.049 | Pass | 0.097 | 0.096 | +0.004 | 0.962 |
| 2 | 20-year | Network_Consistency | 19,780 | 0.049 | Pass | 0.125 | 0.040 | +0.424 | <0.001 |
| 2 | 20-year | Direct_Perc | 19,780 | 0.049 | Pass | 0.332 | 0.488 | -0.447 | <0.001 |

| Lag t | Window | Outcome | Matched pairs | Max \|SMD\| | Balance | Treated/high mean | Control/low mean | Outcome SMD | p-value (clustered) |
|---|---|---|---|---|---|---|---|---|---|
| | | ent | | | | | | | |
| 2 | 20-year | PeerCount_log | 21,655 | 0.049 | Pass | 0.792 | 1.390 | -0.598 | <0.001 |
| 3 | 5-year | Topic_Continuity | 24,419 | 0.034 | Pass | 0.180 | 0.094 | +0.628 | <0.001 |
| 3 | 5-year | Network_Consistency | 51,994 | 0.034 | Pass | 0.119 | 0.022 | +0.576 | <0.001 |
| 3 | 5-year | Direct_Percent | 51,994 | 0.034 | Pass | 0.341 | 0.457 | -0.361 | <0.001 |
| 3 | 5-year | PeerCount_log | 53,263 | 0.034 | Pass | 1.066 | 1.489 | -0.389 | <0.001 |
| 3 | 10-year | Topic_Continuity | 20,207 | 0.039 | Pass | 0.152 | 0.094 | +0.441 | <0.001 |
| 3 | 10-year | Network_Consistency | 39,881 | 0.039 | Pass | 0.125 | 0.030 | +0.523 | <0.001 |
| 3 | 10-year | Direct_Percent | 39,881 | 0.039 | Pass | 0.331 | 0.453 | -0.376 | <0.001 |
| 3 | 10-year | PeerCount_log | 41,476 | 0.039 | Pass | 1.043 | 1.558 | -0.474 | <0.001 |
| 3 | 20-year | Topic_Continuity | 9,068 | 0.051 | Pass | 0.102 | 0.084 | +0.164 | 0.036 |
| 3 | 20-year | Network_Consistency | 18,573 | 0.051 | Pass | 0.110 | 0.033 | +0.411 | <0.001 |

| Lag t | Window | Outcome | Matched pairs | Max \|SMD\| | Balance | Treated/high mean | Control/low mean | Outcome SMD | p-value (clustered) |
|---|---|---|---|---|---|---|---|---|---|
| 3 | 20-year | Direct_Percent | 18,573 | 0.051 | Pass | 0.298 | 0.460 | -0.490 | <0.001 |
| 3 | 20-year | PeerCount_log | 19,968 | 0.051 | Pass | 0.833 | 1.402 | -0.556 | <0.001 |
| 1 | Overall career | Topic_Continuity | 29,560 | 0.028 | Pass | 0.120 | 0.099 | +0.127 | 0.010 |
| 1 | Overall career | Network_Consistency | 58,390 | 0.028 | Pass | 0.184 | 0.036 | +0.695 | <0.001 |
| 1 | Overall career | Direct_Percent | 58,390 | 0.028 | Pass | 0.406 | 0.511 | -0.302 | <0.001 |
| 1 | Overall career | PeerCount_log | 62,964 | 0.028 | Pass | 0.927 | 1.263 | -0.333 | <0.001 |
| 2 | Overall career | Topic_Continuity | 24,671 | 0.026 | Pass | 0.124 | 0.096 | +0.197 | <0.001 |
| 2 | Overall career | Network_Consistency | 59,536 | 0.026 | Pass | 0.156 | 0.022 | +0.702 | <0.001 |
| 2 | Overall career | Direct_Percent | 59,536 | 0.026 | Pass | 0.377 | 0.506 | -0.383 | <0.001 |
| 2 | Overall career | PeerCount_log | 62,762 | 0.026 | Pass | 0.994 | 1.346 | -0.339 | <0.001 |
| 3 | Overall career | Topic_Continuity | 23,627 | 0.022 | Pass | 0.131 | 0.093 | +0.286 | <0.001 |
| 3 | Overall | Network_C | 59,632 | 0.022 | Pass | 0.135 | 0.019 | +0.665 | <0.001 |

| Lag t | Window | Outcome | Matched pairs | Max \|SMD\| | Balance | Treated/high mean | Control/low mean | Outcome SMD | p-value (clustered) |
|---|---|---|---|---|---|---|---|---|---|
| | career | onsistency | | | | | | | |
| 3 | Overall career | Direct_Percent | 59,632 | 0.022 | Pass | 0.342 | 0.486 | -0.450 | <0.001 |
| 3 | Overall career | PeerCount_log | 61,993 | 0.022 | Pass | 1.049 | 1.350 | -0.285 | <0.001 |

Note. Matched differences are reported consistently as treated minus control, using MenteeID-clustered p-values.

**Table S10 Publication-lag robustness of PSM2 matched output comparisons.**

| Lag t | Window | Outcome | Matched pairs | Max \|SMD\| | Balance | Treated/high mean | Control/low mean | Mean difference | p-value (clustered) |
|---|---|---|---|---|---|---|---|---|---|
| 1 | 5-year | Topic_Continuity | 27,544 | 0.010 | Pass | 2.385 | 2.366 | +0.019 | 0.649 |
| 1 | 5-year | Network_Consistency | 24,799 | 0.012 | Pass | 2.663 | 2.414 | +0.248 | <0.001 |
| 1 | 5-year | Direct_Percent | 27,591 | 0.006 | Pass | 1.869 | 2.018 | -0.149 | <0.001 |
| 1 | 5-year | PeerCount_log | 26,367 | 0.006 | Pass | 2.096 | 2.285 | -0.189 | <0.001 |
| 1 | 10-year | Topic_Continuity | 22,592 | 0.014 | Pass | 2.494 | 2.972 | -0.478 | <0.001 |
| 1 | 10-year | Network_C | 19,005 | 0.013 | Pass | 2.963 | 2.641 | +0.322 | <0.001 |

| Lag t | Window | Outcome | Matched pairs | Max \|SMD\| | Balance | Treated/high mean | Control/low mean | Mean difference | p-value (clustered) |
|---|---|---|---|---|---|---|---|---|---|
| | | onsistency | | | | | | | |
| 1 | 10-year | Direct_Percent | 22,371 | 0.007 | Pass | 2.130 | 2.268 | -0.137 | <0.001 |
| 1 | 10-year | PeerCount_log | 21,510 | 0.007 | Pass | 2.258 | 2.525 | -0.267 | <0.001 |
| 1 | 20-year | Topic_Continuity | 11,108 | 0.018 | Pass | 2.696 | 3.987 | -1.291 | <0.001 |
| 1 | 20-year | Network_Consistency | 7,325 | 0.018 | Pass | 3.667 | 3.272 | +0.395 | <0.001 |
| 1 | 20-year | Direct_Percent | 10,759 | 0.015 | Pass | 2.745 | 2.739 | +0.006 | 0.910 |
| 1 | 20-year | PeerCount_log | 8,704 | 0.007 | Pass | 2.541 | 3.023 | -0.482 | <0.001 |
| 2 | 5-year | Topic_Continuity | 27,293 | 0.010 | Pass | 2.600 | 2.438 | +0.163 | <0.001 |
| 2 | 5-year | Network_Consistency | 22,558 | 0.013 | Pass | 2.933 | 2.636 | +0.297 | <0.001 |
| 2 | 5-year | Direct_Percent | 27,210 | 0.007 | Pass | 2.022 | 2.204 | -0.182 | <0.001 |
| 2 | 5-year | PeerCount_log | 26,994 | 0.006 | Pass | 2.269 | 2.436 | -0.167 | <0.001 |
| 2 | 10-year | Topic_Continuity | 21,124 | 0.013 | Pass | 2.725 | 3.041 | -0.316 | <0.001 |

| Lag t | Window | Outcome | Matched pairs | Max \|SMD\| | Balance | Treated/high mean | Control/low mean | Mean difference | p-value (clustered) |
|---|---|---|---|---|---|---|---|---|---|
| 2 | 10-year | Network_Consistency | 16,851 | 0.014 | Pass | 3.219 | 2.853 | +0.366 | <0.001 |
| 2 | 10-year | Direct_Percent | 20,640 | 0.008 | Pass | 2.246 | 2.375 | -0.129 | 0.002 |
| 2 | 10-year | PeerCount_log | 21,257 | 0.008 | Pass | 2.399 | 2.667 | -0.267 | <0.001 |
| 2 | 20-year | Topic_Continuity | 10,310 | 0.018 | Pass | 2.852 | 3.961 | -1.109 | <0.001 |
| 2 | 20-year | Network_Consistency | 6,487 | 0.019 | Pass | 3.899 | 3.411 | +0.489 | <0.001 |
| 2 | 20-year | Direct_Percent | 10,410 | 0.015 | Pass | 2.804 | 2.757 | +0.047 | 0.430 |
| 2 | 20-year | PeerCount_log | 8,583 | 0.007 | Pass | 2.641 | 3.064 | -0.423 | <0.001 |
| 3 | 5-year | Topic_Continuity | 24,312 | 0.012 | Pass | 2.906 | 2.558 | +0.348 | <0.001 |
| 3 | 5-year | Network_Consistency | 20,161 | 0.011 | Pass | 3.192 | 2.891 | +0.301 | <0.001 |
| 3 | 5-year | Direct_Percent | 26,374 | 0.007 | Pass | 2.110 | 2.319 | -0.210 | <0.001 |
| 3 | 5-year | PeerCount_log | 26,947 | 0.008 | Pass | 2.418 | 2.539 | -0.121 | 0.003 |
| 3 | 10-year | Topic_Conti | 20,607 | 0.014 | Pass | 2.901 | 3.190 | -0.289 | <0.001 |

| Lag t | Window | Outcome | Matched pairs | Max \|SMD\| | Balance | Treated/high mean | Control/low mean | Mean difference | p-value (clustered) |
|---|---|---|---|---|---|---|---|---|---|
| | | nuity | | | | | | | |
| 3 | 10-year | Network_Consistency | 14,985 | 0.013 | Pass | 3.461 | 3.125 | +0.336 | <0.001 |
| 3 | 10-year | Direct_Percent | 20,136 | 0.009 | Pass | 2.343 | 2.529 | -0.186 | <0.001 |
| 3 | 10-year | PeerCount_log | 20,734 | 0.008 | Pass | 2.535 | 2.816 | -0.281 | <0.001 |
| 3 | 20-year | Topic_Continuity | 9,906 | 0.017 | Pass | 3.007 | 4.040 | -1.033 | <0.001 |
| 3 | 20-year | Network_Consistency | 5,880 | 0.016 | Pass | 4.079 | 3.572 | +0.507 | <0.001 |
| 3 | 20-year | Direct_Percent | 9,962 | 0.018 | Pass | 2.871 | 2.951 | -0.080 | 0.221 |
| 3 | 20-year | PeerCount_log | 8,412 | 0.009 | Pass | 2.748 | 3.237 | -0.490 | <0.001 |

Note. All PSM2 matching sets pass the prespecified balance threshold. The associations are window-sensitive for topic continuity but consistently positive for network consistency and generally negative for direct mentor collaboration and peer count.

**Table S11 Variance inflation factor diagnostics.**

| Variable | Conventional VIF range | Landmark VIF range | Overall maximum | Assessment |
|---|---|---|---|---|
| Topic continuity (TC) | 1.026-1.070 | 1.019-1.020 | 1.070 | Pass (<5) |

| Variable | Conventional VIF range | Landmark VIF range | Overall maximum | Assessment |
|---|---|---|---|---|
| Network consistency (NC) | 1.061-1.110 | 1.059-1.074 | 1.110 | Pass (<5) |
| Direct mentor collaboration (DP) | 1.096-1.155 | 1.108-1.135 | 1.155 | Pass (<5) |
| Log peer count (PC) | 1.083-1.112 | 1.106-1.109 | 1.112 | Pass (<5) |
| Mentor h-index | 1.140-1.210 | 1.185-1.208 | 1.210 | Pass (<5) |
| Mentored-period citation-elite journal rate | 2.250-3.539 | 2.221-3.489 | 3.539 | Pass (<5) |
| Mentored-period publication rate | 2.072-3.651 | 2.148-3.527 | 3.651 | Pass (<5) |
| Mentored-period network size | 1.026-1.153 | 1.037-1.045 | 1.153 | Pass (<5) |

Note. VIFs are computed for the eight substantive predictors in each complete-case analysis sample. The overall maximum is 3.651, below the conventional threshold of 5.

**Table S12 OLS and PPML estimates across post-graduation windows (t3).**

| Window | Model | Variable | Estimate (SE) | p-value | 0.1-unit effect | N |
|---|---|---|---|---|---|---|
| 5-year | OLS-cluster-mentee | Topic_Continuity | +0.266 (0.104) | 0.011 | | 56,070 |
| 5-year | OLS-cluster-mentee | Network_Consistency | +0.911 (0.107) | <0.001 | | 56,070 |
| 5-year | OLS-cluster-mentee | Direct_Percent | -0.550 (0.120) | <0.001 | | 56,070 |
| 5-year | OLS-cluster-mentee | PeerCount_log | -0.099 (0.013) | <0.001 | | 56,070 |
| 5-year | OLS-two-way-cluster | Topic_Continuity | +0.266 (0.106) | 0.012 | | 56,070 |
| 5-year | OLS-two-way- | Network_Consiste | +0.911 (0.107) | <0.001 | | 56,070 |

| Window | Model | Variable | Estimate (SE) | p-value | 0.1-unit effect | N |
|---|---|---|---|---|---|---|
| | cluster | ncy | | | | |
| 5-year | OLS-two-way-cluster | Direct_Percent | -0.550 (0.121) | <0.001 | | 56,070 |
| 5-year | OLS-two-way-cluster | PeerCount_log | -0.099 (0.015) | <0.001 | | 56,070 |
| 5-year | PPML-cluster-mentee | Topic_Continuity | +0.592 (0.039) | <0.001 | +6.1% | 56,070 |
| 5-year | PPML-cluster-mentee | Network_Consistency | +0.691 (0.022) | <0.001 | +7.2% | 56,070 |
| 5-year | PPML-cluster-mentee | Direct_Percent | -0.835 (0.022) | <0.001 | -8.0% | 56,070 |
| 5-year | PPML-cluster-mentee | PeerCount_log | -0.036 (0.008) | <0.001 | -0.4% | 56,070 |
| 10-year | OLS-cluster-mentee | Topic_Continuity | -1.620 (0.195) | <0.001 | | 42,823 |
| 10-year | OLS-cluster-mentee | Network_Consistency | +0.947 (0.117) | <0.001 | | 42,823 |
| 10-year | OLS-cluster-mentee | Direct_Percent | -0.589 (0.117) | <0.001 | | 42,823 |
| 10-year | OLS-cluster-mentee | PeerCount_log | -0.152 (0.017) | <0.001 | | 42,823 |
| 10-year | OLS-two-way-cluster | Topic_Continuity | -1.620 (0.197) | <0.001 | | 42,823 |

| Window | Model | Variable | Estimate (SE) | p-value | 0.1-unit effect | N |
|---|---|---|---|---|---|---|
| 10-year | OLS-two-way-cluster | Network_Consistency | +0.947 (0.117) | <0.001 | | 42,823 |
| 10-year | OLS-two-way-cluster | Direct_Percent | -0.589 (0.118) | <0.001 | | 42,823 |
| 10-year | OLS-two-way-cluster | PeerCount_log | -0.152 (0.020) | <0.001 | | 42,823 |
| 10-year | PPML-cluster-mentee | Topic_Continuity | +0.027 (0.056) | 0.625 | +0.3% | 42,823 |
| 10-year | PPML-cluster-mentee | Network_Consistency | +0.642 (0.023) | <0.001 | +6.6% | 42,823 |
| 10-year | PPML-cluster-mentee | Direct_Percent | -0.677 (0.024) | <0.001 | -6.5% | 42,823 |
| 10-year | PPML-cluster-mentee | PeerCount_log | -0.061 (0.008) | <0.001 | -0.6% | 42,823 |
| 20-year | OLS-cluster-mentee | Topic_Continuity | -5.527 (0.300) | <0.001 | | 20,568 |
| 20-year | OLS-cluster-mentee | Network_Consistency | +1.060 (0.102) | <0.001 | | 20,568 |
| 20-year | OLS-cluster-mentee | Direct_Percent | -0.519 (0.077) | <0.001 | | 20,568 |
| 20-year | OLS-cluster-mentee | PeerCount_log | -0.255 (0.027) | <0.001 | | 20,568 |
| 20-year | OLS-two-way-cluster | Topic_Continuity | -5.527 (0.307) | <0.001 | | 20,568 |

| Window | Model | Variable | Estimate (SE) | p-value | 0.1-unit effect | N |
|---|---|---|---|---|---|---|
| 20-year | OLS-two-way-cluster | Network_Consistency | +1.060 (0.104) | <0.001 | | 20,568 |
| 20-year | OLS-two-way-cluster | Direct_Percent | -0.519 (0.081) | <0.001 | | 20,568 |
| 20-year | OLS-two-way-cluster | PeerCount_log | -0.255 (0.033) | <0.001 | | 20,568 |
| 20-year | PPML-cluster-mentee | Topic_Continuity | -1.907 (0.140) | <0.001 | -17.4% | 20,568 |
| 20-year | PPML-cluster-mentee | Network_Consistency | +0.466 (0.032) | <0.001 | +4.8% | 20,568 |
| 20-year | PPML-cluster-mentee | Direct_Percent | -0.350 (0.029) | <0.001 | -3.4% | 20,568 |
| 20-year | PPML-cluster-mentee | PeerCount_log | -0.085 (0.011) | <0.001 | -0.8% | 20,568 |

Note. All models include the four displayed variables, mentor h-index, mentored-period citation-elite journal rate, mentored-period publication rate, mentored-period network size, and effective-graduation-year fixed effects. OLS-cluster-mentee and PPML cluster standard errors by MenteeID; OLS-two-way-cluster is the sensitivity specification clustering by both MenteeID and MentorID. PPML models the count with log exposure as an offset; the transformed column reports the expected rate change for a 0.1-unit increase.

**Table S13 Two-part model estimates across post-graduation windows (t3).**

| Window | Model | Variable | Estimate (SE) | p-value | 0.1-unit effect | N |
|---|---|---|---|---|---|---|
| 5-year | LPM-any-cluster-mentee | Topic_Continuity | +0.109 (0.012) | <0.001 | +1.09 pp | 56,070 |
| 5-year | LPM-any-cluster-mentee | Network_Consistency | +0.064 (0.005) | <0.001 | +0.64 pp | 56,070 |

| Window | Model | Variable | Estimate (SE) | p-value | 0.1-unit effect | N |
|---|---|---|---|---|---|---|
| 5-year | LPM-any-cluster-mentee | Direct_Percent | -0.012 (0.005) | 0.019 | -0.12 pp | 56,070 |
| 5-year | LPM-any-cluster-mentee | PeerCount_log | -0.006 (0.001) | <0.001 | -0.06 pp | 56,070 |
| 5-year | Gamma-positive-cluster-mentee | Topic_Continuity | +0.167 (0.037) | <0.001 | +1.7% | 52,731 |
| 5-year | Gamma-positive-cluster-mentee | Network_Consistency | +0.497 (0.019) | <0.001 | +5.1% | 52,731 |
| 5-year | Gamma-positive-cluster-mentee | Direct_Percent | -0.415 (0.016) | <0.001 | -4.1% | 52,731 |
| 5-year | Gamma-positive-cluster-mentee | PeerCount_log | -0.047 (0.004) | <0.001 | -0.5% | 52,731 |
| 10-year | LPM-any-cluster-mentee | Topic_Continuity | +0.053 (0.015) | <0.001 | +0.53 pp | 42,823 |
| 10-year | LPM-any-cluster-mentee | Network_Consistency | +0.045 (0.004) | <0.001 | +0.45 pp | 42,823 |
| 10-year | LPM-any-cluster-mentee | Direct_Percent | -0.009 (0.004) | 0.035 | -0.09 pp | 42,823 |
| 10-year | LPM-any-cluster-mentee | PeerCount_log | -0.006 (0.001) | <0.001 | -0.06 pp | 42,823 |
| 10-year | Gamma-positive-cluster-mentee | Topic_Continuity | -0.909 (0.054) | <0.001 | -8.7% | 41,085 |
| 10-year | Gamma-positive-cluster-mentee | Network_Consistency | +0.490 (0.022) | <0.001 | +5.0% | 41,085 |

| Window | Model | Variable | Estimate (SE) | p-value | 0.1-unit effect | N |
|---|---|---|---|---|---|---|
| 10-year | Gamma-positive-cluster-mentee | Direct_Percent | -0.362 (0.019) | <0.001 | -3.6% | 41,085 |
| 10-year | Gamma-positive-cluster-mentee | PeerCount_log | -0.067 (0.005) | <0.001 | -0.7% | 41,085 |
| 20-year | LPM-any-cluster-mentee | Topic_Continuity | -0.032 (0.027) | 0.246 | -0.32 pp | 20,568 |
| 20-year | LPM-any-cluster-mentee | Network_Consistency | +0.030 (0.004) | <0.001 | +0.30 pp | 20,568 |
| 20-year | LPM-any-cluster-mentee | Direct_Percent | -0.008 (0.005) | 0.136 | -0.08 pp | 20,568 |
| 20-year | LPM-any-cluster-mentee | PeerCount_log | -0.006 (0.001) | <0.001 | -0.06 pp | 20,568 |
| 20-year | Gamma-positive-cluster-mentee | Topic_Continuity | -3.141 (0.102) | <0.001 | -27.0% | 19,988 |
| 20-year | Gamma-positive-cluster-mentee | Network_Consistency | +0.415 (0.028) | <0.001 | +4.2% | 19,988 |
| 20-year | Gamma-positive-cluster-mentee | Direct_Percent | -0.235 (0.027) | <0.001 | -2.3% | 19,988 |
| 20-year | Gamma-positive-cluster-mentee | PeerCount_log | -0.104 (0.008) | <0.001 | -1.0% | 19,988 |

Note. Part 1 is a clustered linear probability model for any citation-elite journal publication; its transformed effect is a percentage-point change. Part 2 is a Gamma-log model among positive outcomes; its transformed effect is the expected intensity change. Both are reported for a 0.1-unit increase.

**Table S14 Landmark PPML and PSM2 results.**

Panel A. PPML regression

| Window | Model | Variable | Estimate (SE) | p-value | 0.1-unit effect | N |
|---|---|---|---|---|---|---|
| Years 4-10 | PPML-cluster-mentee | Topic_Continuity | +0.359 (0.040) | <0.001 | +3.7% | 40,769 |
| | PPML-cluster-mentee | Network_Consistency | +0.524 (0.028) | <0.001 | +5.4% | 40,769 |
| | PPML-cluster-mentee | Direct_Percent | -0.627 (0.026) | <0.001 | -6.1% | 40,769 |
| | PPML-cluster-mentee | PeerCount_log | -0.063 (0.009) | <0.001 | -0.6% | 40,769 |
| Years 6-20 | PPML-cluster-mentee | Topic_Continuity | +0.262 (0.056) | <0.001 | +2.7% | 19,319 |
| | PPML-cluster-mentee | Network_Consistency | +0.279 (0.038) | <0.001 | +2.8% | 19,319 |
| | PPML-cluster-mentee | Direct_Percent | -0.281 (0.032) | <0.001 | -2.8% | 19,319 |
| | PPML-cluster-mentee | PeerCount_log | -0.088 (0.011) | <0.001 | -0.9% | 19,319 |

Panel B. PSM2 matched comparisons

| Window | Analysis | Matched pairs | Max \|SMD\| | Balance | Treated/high mean | Control/low mean | Mean difference | p-value (clustered) |
|---|---|---|---|---|---|---|---|---|
| Years 4-10 | Topic_Contin | 18,083 | 0.013 | Pass | 3.492 | 2.982 | +0.510 | <0.001 |

| Window | Analysis | Matched pairs | Max \|SMD\| | Balance | Treated/high mean | Control/low mean | Mean difference | p-value (clustered) |
|---|---|---|---|---|---|---|---|---|
| | uity | | | | | | | |
| | Network_Consistency | 12,134 | 0.013 | Pass | 3.930 | 3.687 | +0.242 | 0.005 |
| | Direct_Percent | 19,685 | 0.010 | Pass | 2.673 | 2.835 | -0.162 | <0.001 |
| | PeerCount_log | 19,611 | 0.007 | Pass | 2.864 | 3.183 | -0.319 | <0.001 |
| Years 6-20 | Topic_Continuity | 9,354 | 0.016 | Pass | 3.839 | 3.288 | +0.551 | <0.001 |
| Years 6-20 | Network_Consistency | 4,469 | 0.014 | Pass | 4.758 | 4.191 | +0.567 | <0.001 |
| Years 6-20 | Direct_Percent | 9,674 | 0.018 | Pass | 3.267 | 3.327 | -0.059 | 0.456 |
| Years 6-20 | PeerCount_log | 8,061 | 0.008 | Pass | 3.129 | 3.688 | -0.559 | <0.001 |

Note. Capabilities are measured in post-graduation years 1-3. Outcomes are measured in the non-overlapping years 4-10 or 6-20. This landmark design strengthens temporal ordering but does not by itself establish causality.

**Table S15  Landmark two-part model results.**

| Window | Model | Variable | Estimate (SE) | p-value | 0.1-unit effect | N |
|---|---|---|---|---|---|---|
| Years 4-10 | LPM-any-cluster-mentee | Topic_Continuity | +0.053 (0.012) | <0.001 | +0.53 pp | 40,769 |
| Years 4-10 | LPM-any-cluster-mentee | Network_Consistency | +0.040 (0.006) | <0.001 | +0.40 pp | 40,769 |

| Window | Model | Variable | Estimate (SE) | p-value | 0.1-unit effect | N |
|---|---|---|---|---|---|---|
| Years 4-10 | LPM-any-cluster-mentee | Direct_Percent | -0.036 (0.006) | <0.001 | -0.36 pp | 40,769 |
| Years 4-10 | LPM-any-cluster-mentee | PeerCount_log | -0.018 (0.001) | <0.001 | -0.18 pp | 40,769 |
| Years 4-10 | Gamma-positive-cluster-mentee | Topic_Continuity | +0.165 (0.039) | <0.001 | +1.7% | 37,896 |
| Years 4-10 | Gamma-positive-cluster-mentee | Network_Consistency | +0.302 (0.028) | <0.001 | +3.1% | 37,896 |
| Years 4-10 | Gamma-positive-cluster-mentee | Direct_Percent | -0.284 (0.021) | <0.001 | -2.8% | 37,896 |
| Years 4-10 | Gamma-positive-cluster-mentee | PeerCount_log | -0.056 (0.006) | <0.001 | -0.6% | 37,896 |
| Years 6-20 | LPM-any-cluster-mentee | Topic_Continuity | +0.022 (0.014) | 0.133 | +0.22 pp | 19,319 |
| Years 6-20 | LPM-any-cluster-mentee | Network_Consistency | +0.018 (0.007) | 0.007 | +0.18 pp | 19,319 |
| Years 6-20 | LPM-any-cluster-mentee | Direct_Percent | -0.013 (0.007) | 0.048 | -0.13 pp | 19,319 |
| Years 6-20 | LPM-any-cluster-mentee | PeerCount_log | -0.016 (0.002) | <0.001 | -0.16 pp | 19,319 |
| Years 6-20 | Gamma-positive-cluster-mentee | Topic_Continuity | +0.062 (0.060) | 0.302 | +0.6% | 18,513 |
| Years 6-20 | Gamma-positive-cluster-mentee | Network_Consistency | +0.227 (0.038) | <0.001 | +2.3% | 18,513 |
| Years 6-20 | Gamma-positive-cluster-mentee | Direct_Percent | -0.176 (0.030) | <0.001 | -1.7% | 18,513 |
| Years 6-20 | Gamma-positive-cluster-mentee | PeerCount_log | -0.094 (0.009) | <0.001 | -0.9% | 18,513 |

Note. The two-part model separates the probability of any citation-elite journal publication from positive-output intensity. Long-horizon network consistency and peer-count results are the most stable across both margins.

**Table S16 Landmark PSM1 matched descriptive comparisons.**

| Window | Analysis | Outcome | Matched pairs | Max \|SMD\| | Balance | Treated/high mean | Control/low mean | Outcome SMD | p-value (clustered) |
|---|---|---|---|---|---|---|---|---|---|
| Years 4-10 | PSM1 | Topic_Continuity | 24,803 | 0.035 | Pass | 0.190 | 0.140 | +0.303 | <0.001 |
| Years 4-10 | PSM1 | Network_Consistency | 24,881 | 0.035 | Pass | 0.105 | 0.076 | +0.126 | <0.001 |
| Years 4-10 | PSM1 | Direct_Percent | 38,824 | 0.035 | Pass | 0.326 | 0.426 | -0.317 | <0.001 |
| Years 4-10 | PSM1 | PeerCount_log | 40,294 | 0.035 | Pass | 1.031 | 1.454 | -0.398 | <0.001 |
| Years 6-20 | PSM1 | Topic_Continuity | 11,549 | 0.053 | Pass | 0.183 | 0.147 | +0.201 | 0.002 |
| Years 6-20 | PSM1 | Network_Consistency | 11,573 | 0.053 | Pass | 0.090 | 0.084 | +0.004 | 0.944 |
| Years 6-20 | PSM1 | Direct_Percent | 19,307 | 0.053 | Pass | 0.292 | 0.421 | -0.398 | <0.001 |
| Years 6-20 | PSM1 | PeerCount_log | 20,614 | 0.053 | Pass | 0.832 | 1.335 | -0.503 | <0.001 |

Note. PSM0 and PSM1 remain matched descriptive comparisons even when applied to the landmark files. All four matching sets pass max |SMD| < 0.10.

**Table S17 Pair-level sample flow and model diagnostics.**

| Design | Window | Raw pairs | Complete-case N | Retention | Unique mentees | Unique mentors | Zero share | Rate skewness | PPML dispersion | Positive-outcome N | PPML converged |
|---|---|---|---|---|---|---|---|---|---|---|---|
| Conventional t1 | 5-year | 105,322 | 58,501 | 55.5% | 40,303 | 29,198 | 5.6% | 25.109 | 11.698 | 55,221 | Yes |
| Conventional t1 | 10-year | 81,857 | 46,394 | 56.7% | 32,549 | 23,788 | 3.5% | 14.364 | 23.451 | 44,783 | Yes |
| Conventional t1 | 20-year | 40,843 | 22,234 | 54.4% | 16,108 | 12,961 | 2.2% | 4.347 | 47.983 | 21,741 | Yes |
| Conventional t2 | 5-year | 101,696 | 57,751 | 56.8% | 39,386 | 28,990 | 5.8% | 24.385 | 15.984 | 54,395 | Yes |

| Design | Window | Raw pairs | Complete-case N | Retention | Unique mentees | Unique mentors | Zero share | Rate skewness | PPML dispersion | Positive-outcome N | PPML converged |
|---|---|---|---|---|---|---|---|---|---|---|---|
| Conventional t2 | 10-year | 76,903 | 44,736 | 58.2% | 31,111 | 23,193 | 3.8% | 15.342 | 25.480 | 43,035 | Yes |
| Conventional t2 | 20-year | 38,263 | 21,440 | 56.0% | 15,448 | 12,615 | 2.6% | 4.497 | 49.848 | 20,890 | Yes |
| Conventional t3 | 5-year | 97,140 | 56,070 | 57.7% | 38,035 | 28,385 | 6.0% | 23.892 | 13.668 | 52,731 | Yes |
| Conventional t3 | 10-year | 72,255 | 42,823 | 59.3% | 29,671 | 22,446 | 4.1% | 15.521 | 27.907 | 41,085 | Yes |
| Conventional t3 | 20-year | 35,895 | 20,568 | 57.3% | 14,784 | 12,246 | 2.8% | 4.679 | 51.273 | 19,988 | Yes |
| Landmark | Years 4-10 | 72,255 | 40,769 | 56.4% | 28,292 | 21,758 | 7.0% | 15.426 | 24.419 | 37,896 | Yes |
| Landmark | Years 6-20 | 35,895 | 19,319 | 53.8% | 13,889 | 11,739 | 4.2% | 4.994 | 51.686 | 18,513 | Yes |

Note. The observed zero shares, right skew, and overdispersion motivate PPML and the two-part robustness models. Raw pairs and complete-case N count PairIDs; unique mentees and mentors are reported separately. All reported PPML models converged.